\documentclass[12pt]{article}
\input epsf
\usepackage{epsfig}
\def\mod{\mathrm{mod}}
\def\id{\mathrm{id}}
\def\C{\mathbf{C}}
\def\bC{\mathbf{\overline{C}}}
\def\R{\mathbf{R}}
\begin{document}
\title{Analytic continuation of eigenvalues of
a quartic oscillator}
\date{February 11, 2008}
\author{Alexandre Eremenko\thanks{Supported by NSF grant
DMS-0555279.}$\;$ and Andrei Gabrielov}
\maketitle
\begin{abstract}
We consider the Schr\"odinger operator on the real line
with even quartic potential $x^4+\alpha x^2$ and study
analytic continuation of eigenvalues, as functions of
parameter $\alpha$. We
prove several properties of
this analytic continuation conjectured by
Bender, Wu, Loeffel and Martin. 1. All eigenvalues are given by
branches of two multi-valued analytic functions,
one for even eigenfunctions
and one for odd ones.
2. The only singularities of these multi-valued
functions in the complex $\alpha$-plane
are algebraic ramification points, and there are only
finitely many singularities over each compact subset of the $\alpha$-plane.
\end{abstract}

\noindent
{\bf 1. Introduction.}
\vspace{.1in}

Consider the boundary value problem on the real line:
\begin{equation}
\label{0}
-y^{\prime\prime}+(\beta x^4+x^2)y=\lambda y,\quad
y(-\infty)=y(\infty)=0.
\end{equation}
If $\beta>0$, then this problem is self-adjoint,
it has a discrete spectrum of the form
$\lambda_0<\lambda_1<\ldots\to+\infty$, and every
eigenspace is one-dimensional.
The eigenvalues $\lambda_n$ are real analytic functions
of $\beta$ defined on the positive ray.

In 1969,
Bender and Wu \cite{Bender} studied analytic continuation
of $\lambda_n$ to the complex $\beta$-plane.
Their main discoveries are the following:
\vspace{.1in}

\noindent
(i) For every non-negative integers $m$ and $n$ of the
same parity,
the function
$\lambda_m$ can be obtained by an
analytic continuation
of the function $\lambda_n$
along some path in the complex $\beta$-plane.
\vspace{.1in}

\noindent
(ii) The only singularities encountered in the
analytic continuation of $\lambda_n$
in the punctured
$\beta$-plane
$\C\backslash\{0\}$ are algebraic ramification points.
\vspace{.1in}

\noindent
(iii) These ramification points accumulate to $\beta=0$,
in such a way that no analytic continuation
of any $\lambda_n$ to $0$ is possible.
\vspace{.1in}

The last statement gives a good reason why the
formal perturbation series of $\lambda_n$
in powers of $\beta$ is divergent. Bender and Wu also
studied the global structure of the Riemann surfaces
of the functions $\lambda_n$ spread over the
$\beta$-plane, that is the position of their
ramification points, and how the sheets of these Riemann
surfaces are connected at these points.

Bender and Wu used a combination of
mathematical and heuristic arguments with
numerical computation. Since the publication of their
paper, several results about analytic continuation
of $\lambda_n$
were proved rigorously.  We state some of these results.

First of all, we recall a change of
the variable which is credited to Symanzik in \cite{Simon1}.
Consider the family of differential
equations
$$H(\alpha,\beta)y=\lambda y,\quad\mbox{where}\quad
H(\alpha,\beta)=-d^2/dx^2+(\beta x^4+\alpha x^2).$$
The change of the independent variable
\begin{equation}
\label{change}
w(x)=y(tx)
\end{equation}
gives the
differential equation $H(t^4\alpha,t^6\beta)w=
t^2\lambda w$.
Thus, if $\alpha$ and $t$ are real, and $\beta>0$, we have
$$\lambda_n(t^4\alpha,t^6\beta)=
t^2\lambda_n(\alpha,\beta).$$
If $\alpha=1$ we can take $t=\beta^{-1/6}>0$
and
obtain
\begin{equation}\label{bis}
\lambda_n(1,\beta)=\beta^{1/3}\lambda_n(\beta^{-2/3},1).
\end{equation}
This reduces our problem to the study
of the analytic continuation of eigenvalues
of the one-parametric family
\begin{equation}\label{1}
-y^{\prime\prime}+(x^4+\alpha x^2)y= \lambda y,\quad
y(-\infty)=y(\infty)=0,
\end{equation}
depending on the complex parameter $\alpha$.
The study of this family of quartic oscillators
is equivalent to the study of the family (\ref{0}).
Indeed, if we know an analytic continuation of
$\lambda_n(\alpha,1)$
along some curve in the $\alpha$-plane, then equation
(\ref{bis}) gives an analytic continuation of $\lambda_n(1,\beta)$
in the $\beta$-plane and vice versa.
The main advantage of restating the problem in the
form (\ref{1}) is that
any analytic continuation of an eigenvalue (eigenfunction)
of (\ref{1}) remains an eigenvalue (eigenfunction)
of (\ref{1}).
This is not so
for the operator in (\ref{0}): when we perform an analytic
continuation of an eigenfunction the result may no longer
be an eigenfunction, because it may fail to satisfy
the boundary condition, see \cite{BT}. For example, for negative
$\beta$, problem (\ref{0}) has no eigenvalues at all.
\vspace{.1in}

{\em From now on we consider only the family (\ref{1}), and slightly
change our notation:
the $\lambda_n$ will be real analytic functions of $\alpha>0$
representing the eigenvalues of (\ref{1}).}
\vspace{.1in}

Notice that the
$\lambda_n$ have immediate analytic
continuations\footnote{We say that
a function analytic on a set $X\subset\bC$ has an immediate
analytic continuation to a set $Y\subset\bC$ if
$X\cap Y$ has limit points in $X$ and
there exists an analytic function $g$ on $Y$
such that $f(z)=g(z)$ for $z\in X\cap Y$.}
from the positive ray to the whole real line.

Loeffel and Martin \cite{Martin} proved that the functions
$\lambda_n$ have immediate analytic continuations
to the sector
$|\arg\alpha|<2\pi/3$.
They conjectured that the radius
of convergence of the power series of $\lambda_n$ at
$\alpha=0$ tends to infinity as $n\to\infty$.

Simon \cite{Simon1,Simon2} proved that the singularities of
the
$\lambda_n$
accumulate to $\infty$ in the asymptotic direction
of the negative ray. More precisely, for every $n$ and
for every
$\eta\in(0,\pi)$,
there exists $B=B_n(\eta)$ such that $\lambda_n$ has an
immediate analytic
continuation from the positive ray
to the region $\{ \alpha:|\arg\alpha|<\eta,
|\alpha|>B\}.$ On the other hand, he also proved that
the $\lambda_n$ do not have immediate
analytic
continuations to
full punctured neighborhoods of $\infty$.
This proves statement (iii) of Bender and Wu and
implies divergence of the perturbation series for
$\lambda_n$ at $\beta=0$.

Delabaere, Dillinger and Pham
in their interesting papers
\cite{Pham1,Pham2,Pham3} used
a version of the WKB method to study operator (\ref{1})
for large $\alpha$. They claim to confirm
all conclusions of Bender
and Wu, however it is not clear to us what is
proved rigorously in \cite{Pham1,Pham2,Pham3},
which statements
are heuristic and which are verified numerically.
In particular, we could not determine whether these papers
contain
a complete proof of the statement (i).


Another study of $\lambda_n$ for large $\alpha$
is \cite{Gur1}. It is not clear whether statements
of Theorem 1 can be derived from the results of 
\cite{Gur1}. 

In this paper we give complete
proofs of
statements (i) and (ii) of Bender and Wu and of the conjecture
of Loeffel and Martin.
Our methods are different from those of all papers mentioned
above.

\vspace{.1in}

\noindent
{\bf Theorem 1.}{\em

\noindent
{\rm a)} All $\lambda_n$ are branches of two multi-valued analytic
functions $\Lambda^i,\, i=0,1,$ of $\alpha$, one for even $n$
another for odd $n$.
\vspace{.1in}

\noindent
{\rm b)} The only singularities of $\Lambda^i$ over the $\alpha$-plane
are algebraic ramification points.
\vspace{.1in}

\noindent
{\rm c)} For every bounded set $X$ in the $\alpha$-plane, there
are only finitely many ramification points
of $\Lambda^i$ over $X$.}
\vspace{.1in}

Statements a) and b) prove (i) and (ii) of Bender and Wu.
Statement c) implies the conjecture of Loeffel and Martin stated above.
\vspace{.1in}

Statements b) and c) actually hold in greater generality.
Let $P(a,z)=z^d+a_{d-1}z^{d-1}+\ldots+a_1z$ be any monic
polynomial of even degree $d$, with complex coefficients
$a=(a_1,\ldots,a_{d-1})\in\C^{d-1}.$
Consider the boundary value problem
\begin{equation}\label{2}
-y^{\prime\prime}+P(a,z)y=\lambda y,
\end{equation}
\begin{equation}
\label{22}
\quad y(+\infty)
=y(-\infty)=0,
\end{equation}
where the boundary condition is imposed on the real axis.
For non-real $a$, this problem is not self-adjoint, however
it is known \cite{Sibuya} that the spectrum of this problem
is infinite and
discrete, eigenspaces are one-dimensional,
and the eigenvalues tend to infinity
in the asymptotic direction of the positive ray.
Let $\Lambda$ be the multi-valued function of $a\in\C^{d-1}$
which to every $a$ puts into correspondence the set
of eigenvalues of (\ref{2}), (\ref{22}).
We write the set $\Lambda(a)$ as
$\Lambda(a)=\{ \mu_0(a),\mu_1(a),\ldots\}$ where
$$|\mu_0(a)|\leq|\mu_1(a)|\leq\ldots\to+\infty.$$
So for real $a$ we have $\mu_n(a)=\lambda_n(a)$ but
this does not have to be the case for complex $a$.
(The functions $\mu_n$ are not expected to be analytic;
they are only piecewise analytic).
It is known \cite{Sibuya} that there exists an entire function $F$
of $d$ variables with the property that $\lambda$ is an eigenvalue
of the problem (\ref{2}), (\ref{22}) if and only if
\begin{equation}
\label{F}
F(a,\lambda)=0.
\end{equation}
Equation (\ref{F}) may be called the characteristic equation
of the problem (\ref{2}), (\ref{22}). It is the equation
implicitly defining
our multi-valued function $\Lambda$.
\vspace{.1in}

\noindent
{\bf Theorem 2.} {\em The only singularities of $\Lambda$
are algebraic ramification points.
For every $R>0$
there exist a positive integer
$N$ such that
for $n>N$ the $\mu_n$
are single-valued analytic functions
on the set $\{ a:|a|<R\}$ with disjoint graphs.}
\vspace{.1in}

Here are some general properties of
implicit functions
$\lambda(a)$
defined by equations of the form (\ref{F}) with arbitrary entire
function $F$. Let $g_0$ be an analytic
germ of such function at some
point $a_0$, and $\gamma_0:[0,1]\to\C^{d-1}$
a curve in the $a$-space beginning at $a_0$.
Then for every $\epsilon>0$ there is a curve $\gamma$ beginning at
$a_0$, satisfying $|\gamma(t)-\gamma_0(t)|<\epsilon,\; t\in[0,1]$,
and such that an analytic continuation of $g_0$ along $\gamma$ is
possible. In other words, ``the set of singularities'' of
$\Lambda$ is totally discontinuous.
This is called the {\em Iversen property},
and it was proved by Julia \cite{Julia}, see also \cite{Stoilov}.
However, the ``set of singularities'' of $\Lambda$ in general
can have non-isolated points, as can be shown by examples,
\cite{Erem}.

Theorem 2 is in fact an easy consequence of
the following known result.
\vspace{.1in}
\noindent

\noindent
{\bf Theorem A.}
{\em For every $R>0$ there exists a positive integer $N$
such that for all $a$ in the ball $|a|\leq R$
we have the strict inequalities
$|\mu_{n+1}(a)|>|\mu_n(a)|$ for all $n\geq N$.}
\vspace{.1in}

For a complete proof of this result we
refer to Shin \cite[Thm 1.7]{Shin1} who used his earlier
paper \cite{Shin2} and the results
of Sibuya \cite{Sibuya}.
Theorem A is derived from the asymptotic expansion
for
the eigenvalues $\mu_n$ in powers of $n$
which is uniform with respect to $a$
for $|a|\leq R$ \cite[Theorem 1.2]{Shin1}.
Similar asymptotic formula for
the eigenvalue problem (\ref{2}), (\ref{22})
is also given in
\cite[Ch. III, \S 6]{Fedoryuk} where it is derived with a different
method.
\vspace{.1in}

To deduce Theorem 2 from Theorem A, we also need the 
Weierstrass
Preparation Theorem \cite{BM,GR}:
\vspace{.1in}

\noindent
{\bf Theorem B.} {\em Let $Z\subset\C^m$ be the set of
solutions of the equation (\ref{F}), and $(a_0,\lambda_0)\in Z$.
Suppose that $F(a_0,\lambda)\not\equiv 0$.
Then there is a neighborhood $V$ of $(a_0,\lambda_0)$
such that in $V$ we have
$$F(a,\lambda)=\left((\lambda-\lambda_0)^k+F_{k-1}(a)
(\lambda-\lambda_0)^{k-1}+
\ldots+F_0(a)\right)G(a,\lambda),$$
where $F_j$ and $G$ are analytic functions in $V$, and
$G(a_0,\lambda_0)\neq 0$, and $F_j(a_0)=0$ for $0\leq j\leq k-1$.}
\vspace{.1in}

So for each $a$ close to $a_0$ the equation $F(a,\lambda)=0$
with respect to $\lambda$ has $k$ roots close to $\lambda_0$,
and these roots tend to $\lambda_0$ as $a\to a_0$.
\vspace{.1in}

{\em Proof of Theorem 2.}
Application of Theorem A shows that for every $R>0$
there exists $N$ such that for $n>N$ the functions
$\mu_n$ have disjoint graphs over $\{ a:|a|<R\}$.
Application of Theorem B
to the solutions $\mu_n(a)$ of the equation
$F(a,\mu_n(a))=0$ with $n>N$
shows that $k=1$ for all points $(a,\mu_n(a_0))$,
and then
the implicit function theorem implies
that the $\mu_n$ are analytic for $n>N$.
This proves the
second part of Theorem 2. To prove the first part,
consider a curve $\gamma:[0,1]\to\{ a:|a|\leq R\}\subset
\C^{d-1}$ such that $\Lambda$ has an analytic continuation
$g_t$, $0\leq t<1$. Here $g_t$ is an analytic germ of
$\Lambda$ at the point $\gamma(t)$.
Suppose that $g_0(\gamma(0))=\mu_j(\gamma(0)).$
By Theorem A
there exists  $N>j$ such that $|g_t(\gamma(t))|<|
\mu_N(\gamma(t))|$ for all $t\in[0,1)$. As
$\mu_N(a)$ is bounded for $|a|\leq R$, we conclude
that $g_t(\gamma(t))$
is a bounded function on $[0,1)$, and there exists
a sequence $t_k\to 1$ such that $g_{t_k}(\gamma(t_k))$ has
a finite limit $\lambda_1$. Application of the Weierstrass
Preparation theorem to the point $(\gamma(1),\lambda_1)$
shows that in fact $g_t(\gamma(t))\to\lambda_1$ as $t\to 1$,
and $g_t$ either has an analytic
continuation to the point $\gamma(1)$ along $\gamma$
or $\gamma(1)$ is a ramification point of some order $k$.
This completes the proof.
\vspace{.1in}

An alternative proof can be given by using perturbation theory
of linear operators instead of the Weierstrass Preparation
theorem
as it is done in \cite{Simon1}. However we notice
that an analog of Theorem 2 does not hold
for general linear operators analytically dependent
of parameters, as examples in \cite[p. 371-372]{Kato} show.
The crucial property  of our operators (\ref{2}), (\ref{22})
is expressed by Theorem A.

Theorem 2 implies statements b) and c)
of Theorem 1.

In the rest of the paper we prove statement a).
We briefly describe the idea of the proof.
Equation (\ref{F}) which we now write as
\begin{equation}
\label{3}
F(\alpha,\lambda)=0,
\end{equation}
defines an analytic set $Z\subset\C^2$ which consists of all
pairs $(\alpha,\lambda)$ for which the problem (\ref{1}) has a solution.
We are going to
show that this set $Z$ consists of exactly two irreducible
components, which are also its connected components.
To do this we introduce a special parametrization
of the set $Z$ by a
(not connected) Riemann surface $G$. As this parametrization
$\Phi:G\to Z$
comes from the work of Nevanlinna \cite{Nev}, we call it the
{\em Nevanlinna parametrization}.

To study the Riemann surface $G$ we introduce a function
$W:G\to\bC$, which has the property that it is unramified
over $\bC\backslash\{0,1,-1,\infty
\}.$ More precisely, this means that
$$W:G\backslash W^{-1}(\{0,1,-1,\infty\})\to\bC\backslash
\{0,1,-1,\infty\}$$
is a covering map.
Then we study the monodromy action on the generic fiber
of this map $W$. Using a description of $G$ and $W$ which
goes back to Nevanlinna, we label the elements
of the fiber by certain combinatorial objects (cell decompositions
of the plane)
and explicitly describe the monodromy action
on this set of cell decompositions.
Our explicit description shows that
there are exactly two equivalence classes of this action,
thus the Riemann surface $G$ consists of exactly two components.

In fact we not only prove that $G$ consists of two components
but give in some sense a global topological
description of the surface $G$, and thus of the set $Z$.
\vspace{.1in}

The plan of the paper is the following.
In Sections 2 we collect all necessary preliminaries
and construct $G$, $\Phi$ and $W$.
In Sections 3 and 4 we discuss the cell decompositions
of the plane needed in the study
of the monodromy of the map
$W:G\to\bC$. In Section 5 we compute this monodromy
and  complete the proof of
statement a) in Theorem 1. 
In section 6 we briefly mention several other one-parametric
families of linear differential operators with
polynomial potentials which can be treated with the same
method.
\vspace{.2in}

\noindent
{\bf 2. Preliminaries.}
\vspace{.1in}

Some parts of our construction apply to the general problem
(\ref{2}), (\ref{22}) so we explain them for this general case.
We include more detail than it is strictly necessary for
our purposes because the papers \cite{Nev} and \cite{FNev}
are less known nowadays than they deserve.

First we recall some properties of solutions of
the differential equation (\ref{2}). The proofs of
all these properties can be found in Sibuya's book \cite{Sibuya}.
Every solution of this differential equation
is an entire function of order $(d+2)/2$, where $d$
is the degree of $P$. To avoid trivial exceptional
cases, we always assume that $d>0$.
We set $q=d+2$ and divide the plane into $q$ disjoint open sectors
$$S_j=\{ z:|\arg z-2\pi j/q|<\pi/q\},\quad j=0,\ldots, q-1.$$
In what follows we will always understand the subscript $j$
as a residue modulo $q$, so that, for example, $S_q=S_0$ etc.
We call $S_j$ the {\em Stokes sectors} of the equation (\ref{2}).
\vspace{.1in}

\noindent
1. For each solution $y\neq 0$ of the equation (\ref{2}) and each
sector $S_j$ we have either $y(z)\to 0$ or $y(z)\to\infty$
as $z\to\infty$ along each ray from the origin in $S_j$.
We say that $y$
is {\em subdominant} in $S_j$
in the first case and {\em dominant}
in $S_j$ the second case.
\vspace{.1in}


\noindent
2. Of any two linearly independent solutions of (\ref{2}), at most one can be
subdominant in a given Stokes sector.
\vspace{.1in}

Let $y_1$ and $y_2$
be two linearly independent solutions, and consider their ratio
$f=y_2/y_1$. Then $f$ is a meromorphic function of order
$q/2$. (The order of a meromorphic function $f$ can be defined
as the minimal number $\rho$ such that $f$ is a ratio of two
entire functions of order at most $\rho$.)
\vspace{.1in}

\noindent
3. For each $S_j$, we have $f(z)\to w_j\in\bC$ as $z\to\infty$
along any ray in $S_j$ starting at the origin.
\vspace{.1in}

\noindent
4. $w_j\neq w_{j+1}$ for all $j\; \mod\, q$.
\vspace{.1in}

\noindent
5. $w_j\in\{0,\infty\}$ if and only if one of the solutions
$y_1,y_2$ is dominant and another is subdominant in $S_j$.
\vspace{.1in}

A curve $\gamma:[0,1)\to\C$ is called an {\em asymptotic curve}
of a meromorphic function $f$ if $\gamma(t)\to\infty$
as $t\to 1$, and $f(\gamma(t))$ has a limit,
finite or infinite, as $t\to 1$. This limit is called an
{\em asymptotic
value} of $f$.
A classical theorem of Hurwitz says that the singularities of the
inverse function $f^{-1}$ are exactly the critical values
and
the asymptotic values of~$f$.

Returning to the function $f=y_2/y_1$, where $y_1$ and $y_2$
are linearly independent solutions of equation (\ref{2}),
we notice that $f$ does not have critical points. Indeed,
all poles of $f$ are simple because $y_1$
can have only simple zeros,
and $f'(z)\neq 0$ because the Wronskian determinant
of $y_1,y_2$ is constant. Thus $f$ has no
critical values, and the only
singularities of $f^{-1}$ are the asymptotic
values of $f$.

Next we describe these asymptotic values and
associated asymptotic curves.
By property 3 above, for
each sector $S_j$,
every ray from the origin in $S_j$ is an asymptotic
curve. Thus all $w_j$ are asymptotic values.
Function $f$ has no other asymptotic values except the $w_j$.

By Hurwitz theorem we conclude that the only singularities
of $f^{-1}$ lie over the points $w_j$.
More precisely,
\begin{equation}
\label{unram}
f:\C\backslash f^{-1}(\{ w_0,\ldots,w_{q-1}\})\to
\bC\backslash\{ w_0,\ldots,w_{q-1}\}
\end{equation}
is an unramified covering.

Let $D_j$ be discs centered at $w_j$ and having disjoint closures.
Each component $B$ of the preimage $f^{-1}(D_j)$ is either
a topological disc in the plane which is mapped by $f$
onto $D_j$ homeomorphically, or an unbounded domain
such that $f:B\to D_j\backslash\{ w_j\}$ is a universal
covering. Such unbounded domains are called {\em tracts}
over $w_j$.
\vspace{.1in}

\noindent
6. The tracts are in bijective correspondence with the sectors
$S_j$. More precisely, for each $j$ there exists
a unique tract $B_j$ which contains each ray
from the origin in $S_j$, except a bounded subset of this ray,
and the total number of tracts is $q$.
\vspace{.1in}

The {\em Schwarzian derivative} of a function $f$
is
$$S_f=\frac{f^{\prime\prime\prime}}{f'}-\frac{3}{2}\left(
\frac{f^{\prime\prime}}{f'}\right)^2.$$
The main fact about $S_f$ that we need is its relation
with equation (\ref{2}):
\vspace{.1in}

\noindent
7. A ratio $f=y_2/y_1$ of two linearly independent solutions
of (\ref{2}) satisfies the differential equation
\begin{equation}
\label{schwarz}
S_f=-2(P-\lambda),
\end{equation}
and conversely, every non-zero solution of the differential
equation (\ref{schwarz}) is a ratio of two linearly independent
solutions of (\ref{2}).
\vspace{.1in}

Let $Z_d\subset\C^d$ be the set of all pairs $(a,\lambda)$ such that
$\lambda$ is an eigenvalue of the problem (\ref{2}), (\ref{22}).
Consider the class $G_d$ of meromorphic functions with the following
two properties:
\begin{equation}
\label{ss}
-\frac{1}{2}S_f\quad\mbox{is a monic polynomial of degree $d$}
\end{equation}
and
\begin{equation}
\label{zero}
f(z)\to 0, \quad z\in\R,\quad z\to\pm\infty.
\end{equation}
The set $G_d$ is equipped with the usual topology of uniform
convergence on compact subsets of $\C$ with respect to the spherical
metric in the target.

Now we define a map $\Phi:G_d\to\C^d$ by
$\Phi(f)=(a_1,\ldots,a_{d-1},\lambda)$, where
$-\lambda,a_1,\ldots,a_{d-1}$ are the coefficients of
the polynomial $-(1/2)S_f$. This map is evidently continuous.
\vspace{.1in}

\noindent
{\bf Proposition 1.} {\em The map $\Phi$ sends $G_d$ to $Z_d$
surjectively.}
\vspace{.1in}

{\em Proof.}
First we prove that the image of $\Phi$ is contained in $Z_d$.
Let $f$ be an element of $G_d$. By property 7 above, $f=y/y_1$,
a ratio of two linearly independent solutions of (\ref{2}).
By (\ref{zero}) and property 5 above,
$y$ should be subdominant in $S_0$ and $S_{q/2}$.
So $y$ satisfies the boundary condition (\ref{22})
and thus $\Phi(f)=(a,\lambda)$ is an element of $Z_d$.

Now we prove that $\Phi$ maps $G_d$ to $Z_d$ surjectively.
Let $(a,\lambda)\in Z_d$ and let $y$ be the corresponding
eigenfunction. Let $y_1$ be any solution of the equation (\ref{2})
which is linearly independent of $y$. Then $f=y/y_1$
satisfies the differential equation (\ref{schwarz}),
thus (\ref{ss}) holds. Now in view of the boundary
condition (\ref{22}), $y$ is subdominant in $S_0$ and $S_{q/2}$,
so by properties 1 and 2 above $y_1$ must be dominant in $S_0$ and
$S_{q/2}$. So $f=y/y_1$ satisfies (\ref{zero}). Thus $f\in G_d$,
and (\ref{schwarz}) gives $(a,\lambda)=\Phi(f)$.
\vspace{.1in}

\noindent
{\bf Proposition 2.} {\em A meromorphic function $g$
satisfies $g(z)=f(cz)$ for some $f\in G_d$ and $c\in\C^*$
if and only if it has the following three properties:
\vspace{.1in}

\noindent
{\rm (i)} $g$ has no critical points,
\vspace{.1in}

\noindent
{\rm (ii)} $g$ has $q=d+2$ tracts,
\vspace{.1in}

\noindent
{\rm (iii)} There is a tract $B_0$ such that if the
tracts are ordered counterclockwise
as $B_0,B_1,\ldots,B_{q-1}$, and if $w_j$ are
the asymptotic values of $g$ in $B_j$ then
$w_0=w_{q/2}=0$.}
\vspace{.1in}

The sufficiency of these conditions
is a deep result of R. Nevanlinna \cite{Nev}.
The proof becomes much simpler if one adds the condition
\vspace{.1in}

\noindent
(iv) $g$ {\em is a meromorphic function of finite order.}
\vspace{.1in}

This additional condition will be easy to verify in our setting
and we sketch a simpler proof of Proposition 2,
using the condition (iv). This proof is based on F. Nevanlinna's
work \cite{FNev}.
\vspace{.1in}

{\em Proof of Proposition 2 with condition {\rm (iv)}.}
The necessity of conditions (i)-(iv) has already been established.

Now we prove sufficiency.
Condition (i) implies
that $S_g$ is an entire function (in general,
Schwarzian derivative of a meromorphic function has poles exactly
at its critical points). Condition (iv) combined
with the Lemma on the logarithmic derivative \cite{Nevbook,GO}
gives a growth estimate of $S_g$ which implies that
$S_g$ is a polynomial. Now by property 7 above, $f=y/y_1$,
where $y,y_1$ are two linearly independent solutions of the
differential equation
$$y^{\prime\prime}+\frac{1}{2}S_g y=0.$$
Property 6 above shows that $f$ has $\deg S_g+2$ tracts, so
we conclude from (ii) that $\deg S_g=d.$
Now we can find $c\in\C^*$ so that for $f(z)=g(z/c)$
the polynomial $(-1/2)S_f$ is monic. Such $c$ is defined up
to multiplication by a $q$-th root of unity.
Using (iii), we choose this root of unity in such a way that
(\ref{zero}) is satisfied.
\vspace{.1in}

Now we define a map $W:G_d\to\C^q$,
$W(f)=(w_0,\ldots,w_{q-1})$,
whose image is evidently contained in the subspace
$H$ of codimension $2$ given by the equations
$w_0=w_{q/2}=0$.
This map is known to be a local homeomorphism into $H$
\cite{Bakken},
and its image can be described precisely
using a result of R.~Nevanlinna \cite{Nev}.
We do not use these results in our paper. We only remark
that the local homeomorphism $W$ permits to define
a structure of a complex analytic manifold of
dimension $d$ on $G_d$, so that $W$ becomes holomorphic.
The map $\Phi$ we introduced earlier is also holomorphic
with respect to this analytic structure.
\vspace{.1in}

\noindent{\bf Proposition 3.} {\em Let $f_0$ be an element
of $G_d$, and $\gamma:[0,1]\to\C^d$,
$$\gamma(t)=(w_0(t),\ldots,w_{q-1}(t))$$
be a path with the properties $\gamma(0)=W(f_0)$, and
$$w_j(0)\neq w_k(0)\quad\Longleftrightarrow\quad w_j(t)\neq w_k(t)$$
for all $j\neq k$
and all $t\in[0,1]$.
Then there is a lift of the path $\gamma$ to $G_d$,
that is a continuous family $f_t\in G_d,\; t\in[0,1]$
such that $W(f_t)=\gamma(t)$.}
\vspace{.1in}

{\em Proof.} There is a continuous family of diffeomorphisms
$\psi_t:\bC\to\bC,\; \psi_0=\id,$
such that $\psi_t(w_j(0))=w_j(t),\; 0\leq j\leq q-1$.
These diffeomorphisms are quasiconformal \cite{Ahlfors}.
Then the Fundamental existence theorem for quasiconformal
maps \cite[Chap. V]{Ahlfors} implies the existence
of a continuous family of quasiconformal maps $\phi_t,\;
\phi_0=\id,$
such that $g_t=\psi_t\circ f_0\circ\phi_t$ are meromorphic
functions.
These meromorphic functions evidently have no critical
points, because $g_0$ does not.
They have the same number of tracts
as $g_0$ and their tracts satisfy the condition (iii) of
Proposition~2. Thus all conditions of Proposition 2
are satisfied. We can also check the additional condition
(iv): it follows from the general property of
quasiconformal mappings $|\phi_t(z)|\leq |z|^C, |z|>r_0$,
where $C$ is a constant.  Thus by Proposition 2,
$g_t=f_t(c_tz)$, $f_t\in G_d$,
where the constants $c_t$ are determined from
the condition that the polynomials $-(1/2)S_{f_t}$ are monic.
Evidently the correspondence $t\mapsto c_t$ is continuous,
and we have $W(f_t)=\gamma(t)$ by construction.
\vspace{.2in}

\noindent
{\em Centrally symmetric case.}
\vspace{.2in}

Suppose now that the polynomial $P$ in (\ref{2}) is even.
We write it as $P(a,z)=z^d+a_{d-2}z^{d-2}+\ldots+a_2z^2$
and consider the set $Z^e_d$ of all pairs $(a,\lambda)\in\C^{d/2}$
such that $\lambda$ is an eigenvalue of the problem (\ref{2}),
(\ref{22}).
Then each eigenfunction $y$ of the problem (\ref{2}), (\ref{22})
is either even or odd. Indeed, $y(-z)$ is also an eigenfunction
with the same eigenvalue, so $y(z)=cy(-z)$ because the
eigenspace is one-dimensional. Putting $z=0$ we obtain that either
$c=1$ (so the eigenfunction is even) or $y(0)=0$.
In the latter case, differentiate to obtain $y'(z)=-cy'(-z)$
and put $z=0$ to conclude that $c=-1$, so the eigenfunction is odd.

Equation (\ref{2}) with even $P$ always has even and odd solutions:
to obtain an even solution we solve the Cauchy problem
with the initial conditions $y_1(0)=1, y_1^\prime(0)=0$; to obtain
an odd solution we use the initial conditions
$y_1(0)=0, y_1^\prime(0)=1.$

Let $y$ be an eigenfunction, and $y_1$ a solution of (\ref{2})
of the {\em opposite parity} to $y$. Then $y$ and $y_1$
are linearly independent, and the ratio
$f=y/y_1$ is odd. Let $G^o_d$ be the set of all
odd functions in $G_d$. Then $\Phi$ maps $G_d^o$ to $Z_d^e$
because the Schwarzian derivative of an odd function is even.
Thus we have a centrally symmetric version of Proposition~1:
the map
$$\Phi:G_d^o\to Z_d^e\quad\mbox{is well defined and surjective}.$$
Similarly, Proposition 2 has a centrally symmetric analog:
{\em for an odd
meromorphic function $g$ to be of the form $f(cz)$, where
$f\in G_d^o$,
it is necessary and sufficient that conditions {\rm (i)-(iii)}
(or {\rm (i)-(iv)})
be satisfied.}
Finally, Proposition 3 has a centrally symmetric analog:
\vspace{.1in}

\noindent{\bf Proposition $\mathbf{3^\prime}$.}
{\em Let $f_0$ be an element
of $G_d^o$, and $\gamma:[0,1]\to\C^d$,
$$\gamma(t)=(w_0(t),\ldots,w_{q}(t))$$
be a path with the properties $\gamma(0)=W(f_0)$,
$w_j(t)=-w_{j+q/2}(t)$ and
$$w_j(0)\neq w_k(0)\quad\Longleftrightarrow\quad w_j(t)\neq w_k(t)$$
for all $j\neq k \; \mod\, q$
and all $t\in[0,1]$.
Then there is a lift of the path $\gamma$ to $G_d^o$,
that is a continuous family $f_t\in G_d^o,\; t\in[0,1]$
such that $W(f_t)=\gamma(t)$.}
\vspace{.1in}

The proof is the same as that of the original Proposition 3:
one can choose all homeomorphisms $\psi_t$ and $\phi_t$ to be odd,
then $g_t=\psi_t\circ f_0\circ\phi_t$ will be odd.
\vspace{.1in}

Case a) of Theorem 1 which we are proving corresponds to
the even potential with $d=4$. To prove a) we only need to
show that $G_4^o$ consists of two components:
one containing the functions with $f(0)=0$ and another
containing the functions with $f(0)=\infty$.

Notice that $\C^*$ acts on $G_d$ and on $G_d^o$
by the rule $f\mapsto cf$, and that the map $\Phi$ is invariant
with respect to this action. Introducing the equivalence relation
$f\sim cg,\; c\in\C^*$ on $G_d^o$ we obtain a factor-map $\tilde{\Phi}$
which maps the equivalence classes to $Z_d$.
\vspace{.1in}

\noindent
{\bf Proposition 4.} {\em The map
$\tilde{\Phi}:G_d^o\to Z_d^e$ is a homeomorphism.}
\vspace{.1in}

{\em Proof.} We only have to show that it is injective,
that is that any two non-zero odd solutions $f_1$ and $f_2$
of the Schwarz differential equation
$S_f=-2P,$
which tend to zero as $z\to \pm\infty$ on the real line,
are proportional. All non-zero solutions of
a Schwarz differential equation are related by fractional-linear
transformations. So we have $f_1=T\circ f_2$, where $T$
is a fractional-linear transformation.
Changing $z$ to $-z$ we conclude that $T$ is odd.
Every odd fractional-linear transformation has the form $cz$
or $c/z$. The latter case is excluded by the condition
that $f_1(z)$ and $f_2(z)$ both tend to zero as $z\to\infty$
on the real line.
Thus $f_1=cf_2$.
\vspace{.1in}

In the next section we study the map $W:G_d\to\C^q$,
and in particular, the monodromy of this map.
For this we need a description of the general fiber
of this map by certain cell decompositions of the pane.
\vspace{.2in}

\noindent
{\bf 3. Some cell decompositions of the plane.}
\vspace{.2in}

\def\g{{\gamma}}
\def\G{{\Gamma}}
\def\P{{\mathbf{P}}}
\def\C{{\mathbf{C}}}
\def\R{{\mathbf{R}}}
\def\bR{{\mathbf{\overline{R}}}}
\def\bC{{\mathbf{\overline{C}}}}
\def\rk{{\rm rk}}
\def\crk{{\rm crk}}
By a {\em cell decomposition} of a surface $X$ we understand its
representation as a locally finite union of disjoint subsets
called {\em cells}.
The cells can be of dimension 0 (points), 1 (edges)
or 2 (faces). The edges and faces
are homeomorphic images of an open
interval or of an open disc, respectively, and they satisfy the
following condition: the boundary (in $X$)
of each cell is a locally finite
union of cells of smaller dimension of this decomposition.
We do not assume that the homeomorphisms of the open
discs defining faces have extensions to the closed discs.

Let $w=f(z)$ be a meromorphic function without critical points
and with finitely many asymptotic values.
Consider a fixed cell decomposition $\Psi_0$ of the sphere $\bC_w$
such that all asymptotic values are
contained in the faces and
each face contains one asymptotic value.
Then the preimage $\Psi_f=f^{-1}(\Psi_0)$ is a cell
decomposition of the plane $\C_z$ with connected
$1$-skeleton. That the $1$-skeleton is connected
is seen from (\ref{unram}), which is 
a covering, and from the fact that every path in
$\bC\backslash\{w_0,\ldots,w_{q-1}\}$
can be deformed to a path in the $1$-skeleton of $\Psi_0$.
The closures of the edges of $\Psi_f$
are mapped by $f$ onto the closures of the
edges of $\Psi_0$ homeomorphically.
Each face $B$
of $\Psi_f$ is mapped by $f$ onto a face $D$ of $\Psi_0$
either homeomorphically or as a universal covering
over $D\backslash\{ w\}$ where $w$ is
the asymptotic value in $D$.
In the former case the face $B$ of $\Psi_f$ is bounded,
in the latter case
it is unbounded.

We label the faces of $\Psi_f$ by the names
of their image faces
under $f$. Labeled cell decompositions
of the plane $\C_z$ are considered
up to orientation-preserving
homeomorphisms of the plane preserving the labels.
If $f$ is an odd function, it is reasonable to
choose $\Psi_0$ to be invariant under the map $w\mapsto-w$.
Then $\Psi_f$ will be also invariant under $z\mapsto-z$.
For such cell decompositions
of $\C_z$, the natural equivalence relation is that they are
mapped one onto another by an odd orientation-preserving
homeomorphism of $C_z$ respecting the  face labels.
We call two such cell decompositions symmetrically equivalent.
For a given set of asymptotic values
and a given $\Psi_0$, the labeled cell decomposition
$\Psi_f$ almost completely determines $f$. Namely, we have
the following
\vspace{.1in}

\noindent
{\bf Proposition 5.} {\em Let $f_1$ and $f_2$ be two meromorphic
functions without critical points and with the same finite 
set
of asymptotic
values. Fix a cell decomposition $\Psi_0$ of $\bC_w$
such that the asymptotic values
are contained in faces of $\Psi_0$ and each
face contains one asymptotic
value. If $\Psi_i=f_i^{-1}(\Psi_0)$ are equivalent
cell decompositions of $C_z$
then $f_1(z)=f_2(cz+b),\; c\neq 0$.

If $f_i$ are odd, $\Psi_0$ is centrally symmetric, and
the $\Psi_i$ are symmetrically equivalent then $b=0$.}
\vspace{.1in}

{\em Proof.} Let $\psi$ be the orientation-preserving
homeomorphism of the plane $\C_z$ that maps $\Psi_1$
onto $\Psi_2$ preserving the face labels.
We are going to define another homeomorphism $\psi'$
with the same properties, and in addition,
\begin{equation}
\label{sss}
f_1=f_2\circ\psi'.
\end{equation}
Let $B_1$ and $B_2=\psi(B_1)$ be two faces such that
$f_i$ map $B_i$ onto a face $B_0$ of $\Psi_0$.
If one of the $B_1,B_2$ is bounded then another is
also bounded and the maps $f_i:B_i\to B_0$
are homeomorphisms. So there exists a unique
homeomorphism $\psi':B_1\to B_2$ such that
$(\ref{sss})$ holds. If both $B_1$ and $B_2$ are
unbounded, then $f_i:B_i\to B_0\{ w_0\}$
are universal coverings,
and there are infinitely many homeomorphisms $B_1\to B_2$
that satisfy (\ref{sss}).
To choose one, we first notice that every homeomorphism
$B_1\to B_2$
with property (\ref{sss}) has a continuous extension
to the boundary $\partial B_1$ and sends boundary edges of
$B_1$ to boundary edges of $B_2$. We choose $\psi'$ in $B_1$
so that it maps the boundary edges in the same way as $\psi$.
This is possible to do as $\psi$ preserves orientation.
Now $\psi'$ is defined on all faces of $\Psi_1$,
and it is easy to check that the boundary extensions
from different faces to edges match.
Thus $\psi'$ is a homeomorphism of the plane satisfying
(\ref{sss}), and (\ref{sss}) implies that it is conformal.
So $\psi'(z)=az+b$. It is easy to check that in
the centrally symmetric
case the above construction gives an odd
homeomorphism $\psi'$.
\vspace{.1in}

Now we consider a special class
of cell decompositions $\Psi_0$
which is convenient for our purposes.\footnote{The usual
choice of $\Psi_0$ as in \cite{Drape,GO,Nev,Nevbook}
leads to the cell decompositions of the plane
which are called line complexes. We prefer a different
choice of $\Psi_0$, as in \cite{EGS},
which is better compatible with
the symmetries of our problem.}
\vspace{.1in}

Let $f:\C_z\to\bC_w$ be a meromorphic function of order
$q/2$ without critical points and with the asymptotic values
$w_0,\dots,w_{q-1}$, ordered according to the
cyclic order of their Stokes sectors.
We assume that the set
$J=\{j: w_j\ne 0\}$ is a fixed subset of $\{0,\dots,q-1\}$,
and that all nonzero asymptotic values
of $f$ are finite and distinct.
Let $J=\{j_1,\dots,j_k\}$ with $j_1<\dots<j_k$,
and let $c_\nu=w_{j_\nu}$.
Then the cyclic order of the Stokes sectors
with nonzero asymptotic values $c_1,\dots,c_k$
agrees with that of $\{1,\dots,k\}\,\mod\,k$.
We assume $3\le k<q$, so $f$ has at least
three distinct non-zero asymptotic values.

We define a cell decomposition $\Psi_0$ of $\bC_w$ with a single vertex
at $\infty$ as follows (see Fig.~1).
\begin{center}
\epsfxsize=3.0in%
\centerline{\epsffile{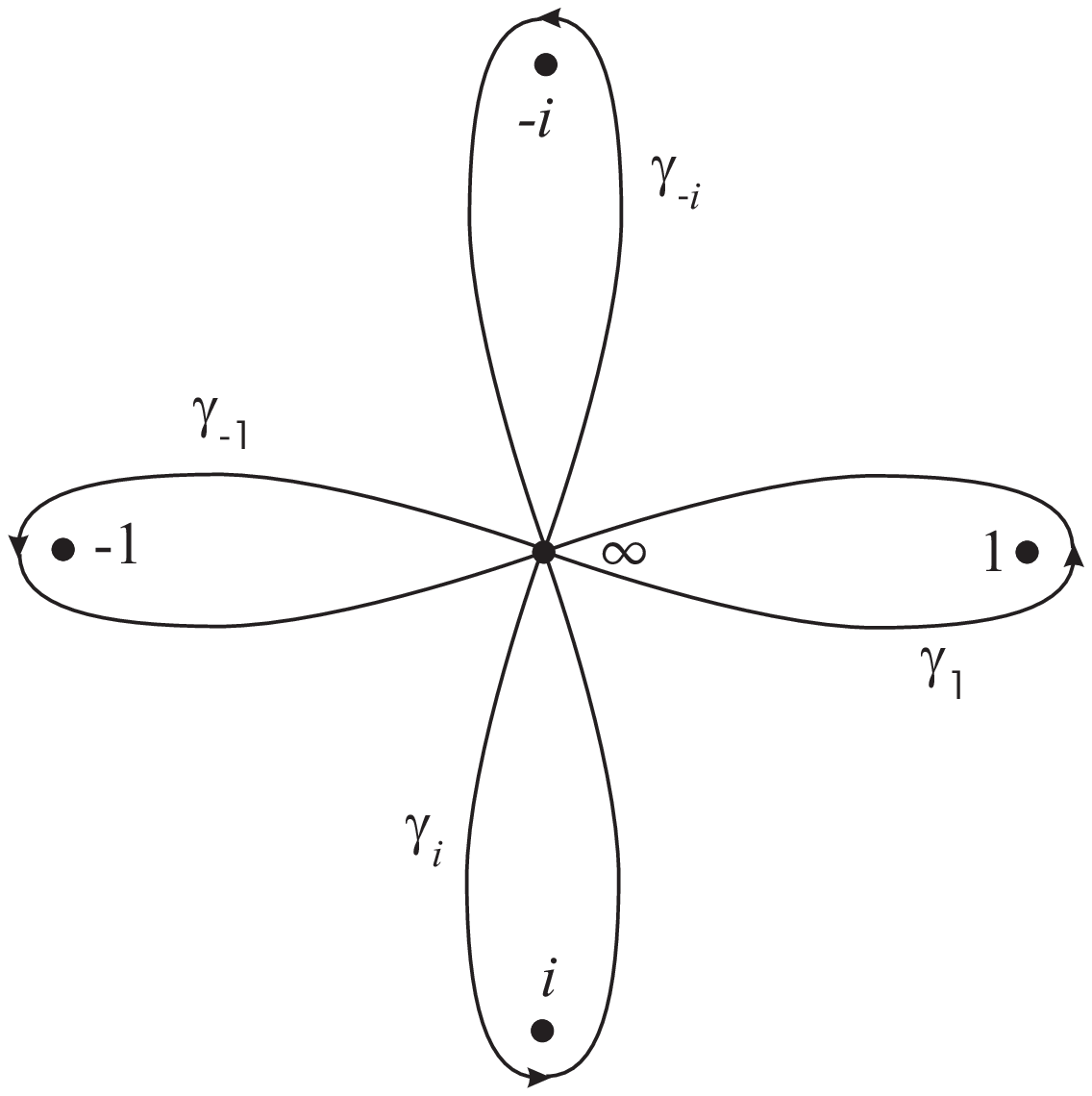}}
\nopagebreak
\vspace{.2in}

Fig.~1. Cell decomposition $\Psi_0$ for asymptotic values
$0,\pm 1,\pm i$.
\end{center}
In the complex plane $\C_w^\bullet =\bC_w\setminus\{0\}$,
fix a system of directed loops $\g_{c_1},\dots,\g_{c_k}$,
starting and ending at $\infty$, intersecting only at their endpoints,
and such that each loop $\g_{c_\nu}$ is an oriented boundary of an open domain
$D_{c_\nu}$ containing $c_\nu$ and not containing other asymptotic values of $f$.
Let $D_0$ be the connected component of $0$
in $\bC_w\setminus(\g_{c_1}\cup\dots\cup\g_{c_k})$.
The domains $D_0,\,D_{c_1},\dots,D_{c_k}$ are the faces of $\Psi_0$,
and the open loops $\dot\g_\nu=\g_\nu\setminus\{\infty\}$ are its edges.

There is a natural cyclic order
$\nu_1\prec\dots\prec\nu_k\prec\nu_1$ of the loops $\g_{c_\nu}$,
and of the corresponding domains $D_{c_\nu}$.
It is defined by the order in which the domains
$D_{c_\nu}$ cross the oriented boundary of a small disk
in $\C_w^\bullet$ centered at $\infty$.
Alternatively, it is opposite to the
order in which the loops $\g_{c_\nu}$ appear
in the boundary of $D_0$.

So we introduced two cyclic orders on $c_1,\ldots,c_k$,
the first one coming from the cyclic order of the Stokes
sectors, and the second one from the cyclic order
of the loops in the cell decomposition $\Psi_0$.
These two cyclic orders are in general different.

Let $\Psi_f=f^{-1}\Psi_0$ be the corresponding cell decomposition of $\C_z$.
Its faces are labeled with $0,\,c_1,\dots,c_k$ and edges with $c_1,\dots,c_k$.
The edges of $\Psi_f$ are directed, being preimages of directed loops $\g_{c_\nu}$.
Since $f$ has no critical points, the restriction
of $f$ to any face of $\Psi_f$ is either a homeomorphism
or a universal covering
over the image face of $\Psi_0$ minus the asymptotic
value in this image face.
Accordingly, $\Psi_f$ has the following properties:

{\bf (1)} Unbounded faces of $\Psi_f$ are in one-to-one correspondence with the Stokes sectors of $f$.
For each $\nu=1,\dots,k$ there is exactly one unbounded face $B_{c_\nu}$ labeled with $c_\nu$.
The faces $B_{c_1},\dots,B_{c_k}$ have the cyclic order
\newline
$\{1,\dots,k\}\,\mod\, k$ at infinity.

{\bf (2)} Edges of $\Psi_f$ may be either {\it links} (having two distinct vertices) or {\it loops}
(having both ends at the same vertex).
Each edge labeled by $c_\nu$ separates two faces, labeled with $c_\nu$ and $0$, respectively.
Its orientation agrees with that of the boundary of its adjacent face labeled with $c_\nu$.
If it is a link, it is adjacent to the unbounded
face $B_{c_\nu}$.

{\bf (3)} Each bounded face of $\Psi_f$ labeled with $c_\nu$ has as its boundary a loop labeled
by $c_\nu$. The unbounded face $B_{c_\nu}$ has as its boundary an infinite chain of links
labeled with $c_\nu$, and no loops.
A bounded face labeled with $0$ has as its boundary $k$ edges labeled with $c_\nu$,
(oriented opposite to their natural orientation)
in the cyclic order opposite to the order $\nu_1,\dots,\nu_k$ of the loops $\g_{c_\nu}$.
An unbounded face labeled with $0$ has as its boundary an infinitely repeated sequence of $k$ edges
labeled with $c_\nu$, in the cyclic order opposite to $\nu_1,\dots,\nu_k$.

{\bf (4)} The cyclic order of the values
$c_\nu$ labeling the {\em links} in the boundary of
a bounded face labeled by $0$ is the same as the
cyclic order
of the unbounded faces $B_{c_\nu}$
adjacent to these links, which agrees with the
cyclic order $\{1,\dots,k\}\, \mod\, k$ of the Stokes
sectors.

{\bf (5)} Each vertex $v$ of $\Psi_f$ has degree $2k$, with the directed edges
labeled with $c_{\nu_1},\dots,c_{\nu_k}$ consecutively exiting and entering $v$,
where $\nu_1,\dots,\nu_k$ is the cyclic order of the loops $\g_{c_\nu}$.
This means that, as an edge labeled with $c_{\nu_1}$ exits $v$, the next
in the cyclic order is (the same or another) edge labeled with
$c_{\nu_1}$ entering $v$, then an edge labeled with $c_{\nu_2}$
exiting $v$, and so on, till an edge labeled with $c_{\nu_k}$ entering $v$,
followed by the initial edge labeled with $c_{\nu_1}$ exiting $v$.
\vspace{.1in}

The one-skeleton of $\Psi_f$
is an infinite 
directed graph properly embedded in $\C_z$.
It is connected.

Removing from the $1$-skeleton of $\Psi_f$ all loops, we obtain a directed graph $\G$.
From property {\bf (3)} of $\Psi_f$, all bounded components of the complement of $\G$ are
labeled by $0$, and the unbounded
components are in one-to-one correspondence with the Stokes sectors of $f$.
Moreover, the unbounded components corresponding to the Stokes sectors with nonzero
asymptotic values are exactly the faces $B_{c_\nu}$ of $\Psi_f$.
Replacing two edges in the boundary of each two-gon of $\G$
by one {\it undirected} edge inside the two-gon connecting its two vertices,
and forgetting orientation of all remaining edges of $\G$,
we obtain a properly embedded graph
$T$ without loops or multiple edges,
with the components of its complement labeled with $0$ and $c_\nu$.
That $\Gamma$ has no multiple edges follows from property
{\bf (2)} of $\Psi_f$: each link in $\Psi_f$ belongs
to the boundary of some unbounded face, thus there are
at most two links in $\Phi_f$ between any pair of vertices.
Each edge of $T$ separates two components with different labels.
All bounded components are labeled with $0$.
\vspace{.1in}

\noindent
{\bf Proposition 6.}
{\em Suppose that the cyclic
 order $\nu_1,\dots,\nu_k$ of
 the loops $\g_{c_\nu}$ in $\C_w^\bullet$ agrees with the
cyclic order $\{1,\dots,k\}\,\mod\,k$ of the Stokes sectors
with the asymptotic values $c_1,\dots,c_k$ in $\C_z$.
Then each bounded face of $\Psi_f$ labeled with
$0$ has exactly two vertices, and its boundary
contains exactly two links.
Each bounded component of the complement of $\G$ is a two-gon,
and $T$ is an embedded planar tree.}
\vspace{.1in}

{\em Proof.}
First, a bounded face $C$ of $\Psi_f$ labeled
with $0$ must have at least two vertices.
Otherwise, its boundary would consist of loops
labeled with $c_\nu\ne 0$, which
should be also boundaries of  bounded faces
labeled with $c_\nu$. This
is impossible since the union of this face
with the loops and the vertices would be a sphere,
and could not be embedded in $\C_z$.
Hence there are at least two links in the boundary of $C$.

From property {\bf (3)} of $\Psi_f$, the cyclic order of the links labeled with $c_\nu$ in the boundary
of $C$ should be opposite to the cyclic order of the loops $\g_{c_\nu}$.
From property {\bf (4)}, it should agree with
the cyclic order $\{1,\dots,k\}\,\mod k$
of the faces $B_{c_\nu}$.
If the loops $\g_{c_\nu}$ and
the faces $B_{c_\nu}$ have the
same cyclic order, this is impossible
when the boundary of $C$ contains more than two links.

Since $\G$ is obtained by removing loops from the
one-skeleton of $\Psi_f$, its complement has no bounded components other
than two-gons, hence the complement of $T$ has no bounded components, so
$T$ is a forest.

To prove that it is connected, we notice that the $1$-skeleton
of $\Psi_f$ is connected, and the removal
 of loops and multiple edges
from $\Psi_f$ does not affect this connectedness.
\vspace{.1in}

\noindent
{\bf Proposition 7.}
{\em The embedded planar directed graph $\G$ and the cell decomposition $\Psi_f$
are determined by the embedded planar graph $T$ uniquely up to an
orientation-preserving homeomorphism of $\C_z$
preserving their common vertices.}
\vspace{.1in}

{\em Proof.}
The components of the complement of $T$ labeled with $0$ coincide with
the components of the complement of $\G$ labeled with $0$.
A unique unbounded component $C_\nu$ of the complement of $T$ labeled by $c_\nu$
contains the unbounded face $B_{c_\nu}$ of $\Psi_f$.
Each of its boundary edges separates $C_\nu$ either from
a component of the complement of $T$ labeled with $0$ or from $C_\mu$ with $\mu\ne\nu$.
In the first case, the edge belongs to $\G$. We make it
directed as part of the boundary of $C_\nu$ and label it with $c_\nu$.
Otherwise, we connect the two vertices of the edge by a new edge labeled with
$c_\nu$ inside $C_\nu$, directed according to the orientation of the boundary of $C_\nu$,
so that all these new edges are disjoint (this can be done one edge at a time, in any order).
As a result, each edge of $T$ separating two components $C_\nu$ and $C_\mu$
is included inside a two-gon.
Let $\G'$ be the embedded planar directed graph obtained by removing
all such edges of $T$.
We want to show that $\G'$ can be obtained from $\G$ by an
orientation-preserving homeomorphism of $\C_z$
preserving all vertices and labels.
First, two-gons of $\G'$ have the same vertices and the same labeling
of edges as the two-gons of $\G$.
Hence there exists an orientation-preserving homeomorphism
between each two-gon of $\G$ and the two-gon of $\G'$ with the same vertices,
preserving their common vertices.
These homeomorphisms define a homeomorphism between the union of all two-gons of $\G$
and the union of all two-gons of $\G'$, preserving their common vertices and the labeling
of their edges. It can be extended to unbounded components to obtain a homeomorphism 
of $\C_z$.

From property {\bf (5)} of $\Psi_f$, the cyclic order
of the components $B_{c_\nu}$ adjacent to a vertex $v$ of $\G$
agrees with the cyclic order $\nu_1,\dots,\nu_k$ of the loops $\g_{c_\nu}$.
The cyclic order of the edges of $\G$ exiting and entering $v$ is
determined by the cyclic order of the unbounded
 components $B_{c_\nu}$
of its complement adjacent to $v$.
Since this order agrees with the cyclic order
of the faces $D_{c_\nu}$ of $\Psi_0$ (same as the order of the loops $\gamma_{c_\nu}$),
one can add to $\G$ non-intersecting loops, having both ends at $v$,
labeled with the missing values $c_\nu$
inside the connected components of the complement of $\G$ labeled by $0$ adjacent to $v$,
so that the cyclic order of the edges consecutively exiting and entering $v$ becomes
$\nu_1,\dots,\nu_k$.
Labeling the interiors of these loops by the corresponding values $c_\nu$,
we obtain a cell decomposition of $\C_z$
that can be obtained from $\Psi_f$ by an
orientation-preserving homeomorphism of $\C_z$
 preserving vertices and labels
(this can be done first for the loops,
 in any order, then extended to components labeled with $0$).
\vspace{.1in}

\noindent
{\em Centrally symmetric case.}
\vspace{.1in}

Suppose now that the set of asymptotic values $w_j$ (and the corresponding set of non-zero values $c_\nu$)
is centrally symmetric. In this case, $k$ is even and $c_{\nu+k/2}=-c_\nu$.
We can choose the loops $\g_{c_\nu}$ centrally symmetric (e.g., by selecting $k/2$ loops about $c_\nu^2$
and taking square roots of them).
If $f$ is odd, the cell decomposition $\Psi_f$ is centrally symmetric,
and so are the graphs $\G$ and $T$ (assuming the edges of $T$ inside centrally
symmetric two-gons of the complement of $\G$ are chosen centrally symmetric).
The origin $0\in\C_z$ is either a vertex of $\Psi_f$ when
$f(0)=\infty$, or the center of a bounded face labeled with $0$ when $f(0)=0$.
This bounded face may correspond either to a bounded face of the complement
of $T$ (if its boundary has more than two links) or to the middle of an edge of $T$.
\vspace{.2in}

\noindent
{\bf 4. Case $q=6,\, k=4$: classification of trees.}
\vspace{.1in}

Consider the centrally symmetric case $q=6,\, k=4$, with $w_0=w_3=0$.
We assume that the Stokes sector $S_0$
contains a ray of the positive real axis.
We have $c_1=w_1=-c_3=-w_4$ and $c_2=w_2=-c_4=-w_5$ satisfying
$\{ c_1,c_2\}\cap\{ 0,\infty\}=\emptyset$ and $c_1\ne\pm c_2$.
Since multiplication of all asymptotic values by a nonzero constant corresponds to multiplication of $f$
by the same constant, we can assume
$c_2=1$ and $c_1=c\ne 0,\,\infty,\,\pm 1$.

Suppose that the cell decomposition $\Psi_0$ is centrally symmetric and satisfies conditions of Proposition 6,
so that $T$ is a centrally symmetric tree. The two
components of its complement labeled by $0$ and
opposite via central symmetry,
cannot have a common boundary edge.
\vspace{.1in}

\noindent
{\bf Proposition 8.} {\em Consider centrally symmetric
embedded trees in $\C$ with six ends,
two opposite components of the complement
labeled by zero, and such that
these two components do not have a common edge.
Any such tree is equivalent to either one
of the trees shown in Fig.~2, or its
complex conjugate.
The integer parameters satisfy the following restrictions:
for $A_k:\; k\geq 0$, for $D_{k,l}:\; k\geq 0, l\geq 1$, and
for $E_{k,l}:\; k\geq 1,l\geq 0$.}
\vspace{.1in}

\begin{center}
\epsfxsize=5.0in%
\centerline{\epsffile{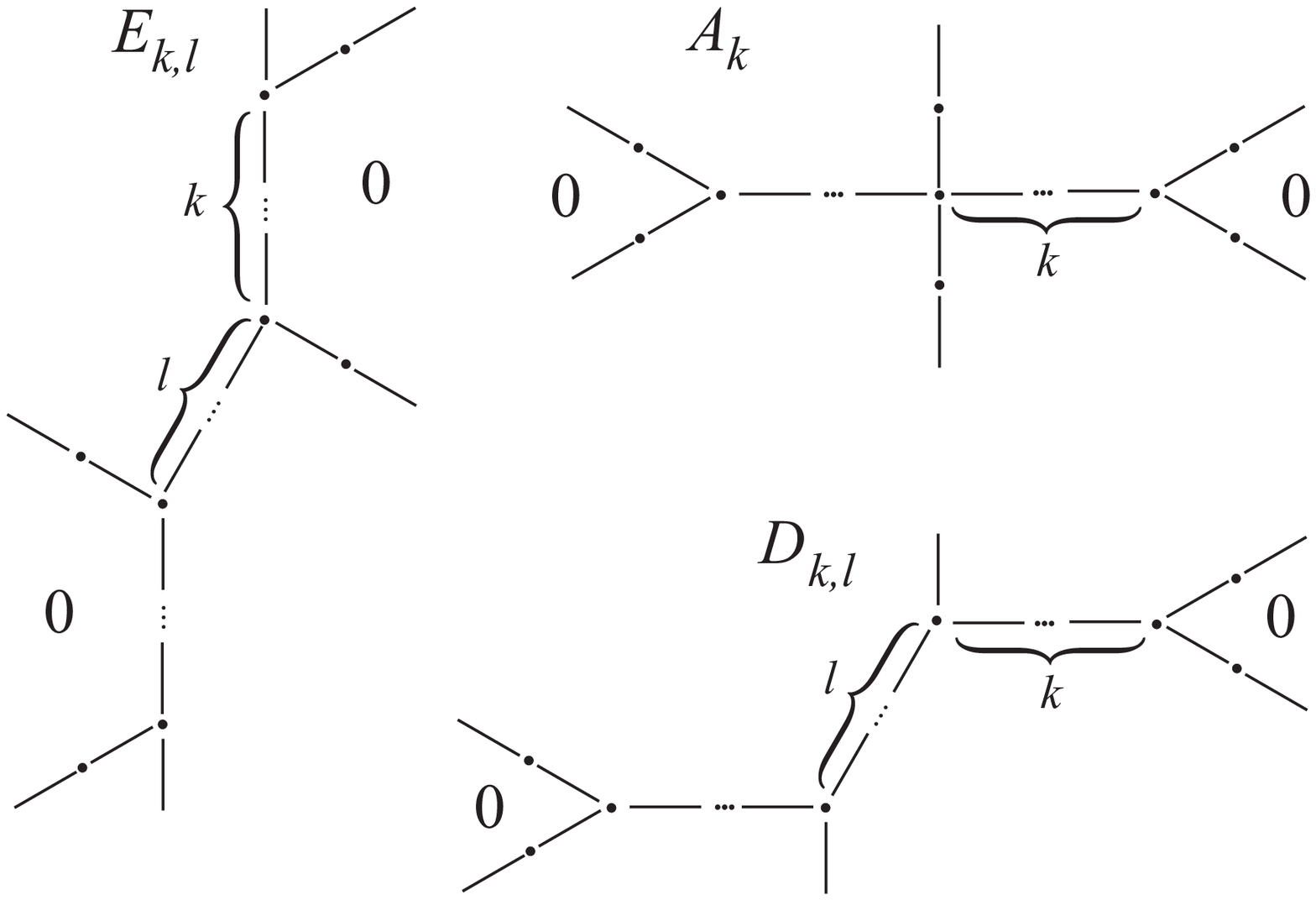}}
\nopagebreak
\vspace{.2in}\nopagebreak
Fig.~2. Classification of trees.
\end{center}

{\em Proof.}
Let $X$ be a graph in $\C_z$ with the
 vertices corresponding to components of the complement of $T$
and the edges connecting two of its vertices if the corresponding components have a common edge in $T$.
Then $X$ is combinatorially a hexagon with some of the chords connecting its vertices, so that
the chords do not intersect and the vertices $0$ and $3$ not connected.
We can also assume that $X$ is centrally symmetric and that its vertex labeled with $0$
is on the positive real axis.
There are ten possible cases.
Six of them are listed in Fig.~3.
They correspond to embedded planar trees
$A_k,\, D_{k,l},\, E_{k,l}$ shown in Fig.~2.
Here indices $k$ and $l$ denote the number of edges
in a chain of edges and
simple vertices of $T$ corresponding
to a chord of $X$ (with $l$ corresponding to
the chord passing through the origin).
In particular, a tree with even $l$ has a vertex at the origin,
while a tree with odd $l$ has the origin as the middle of its edge.
The remaining four cases are obtained by complex conjugation.
\begin{center}
\epsfxsize=4.5in%
\centerline{\epsffile{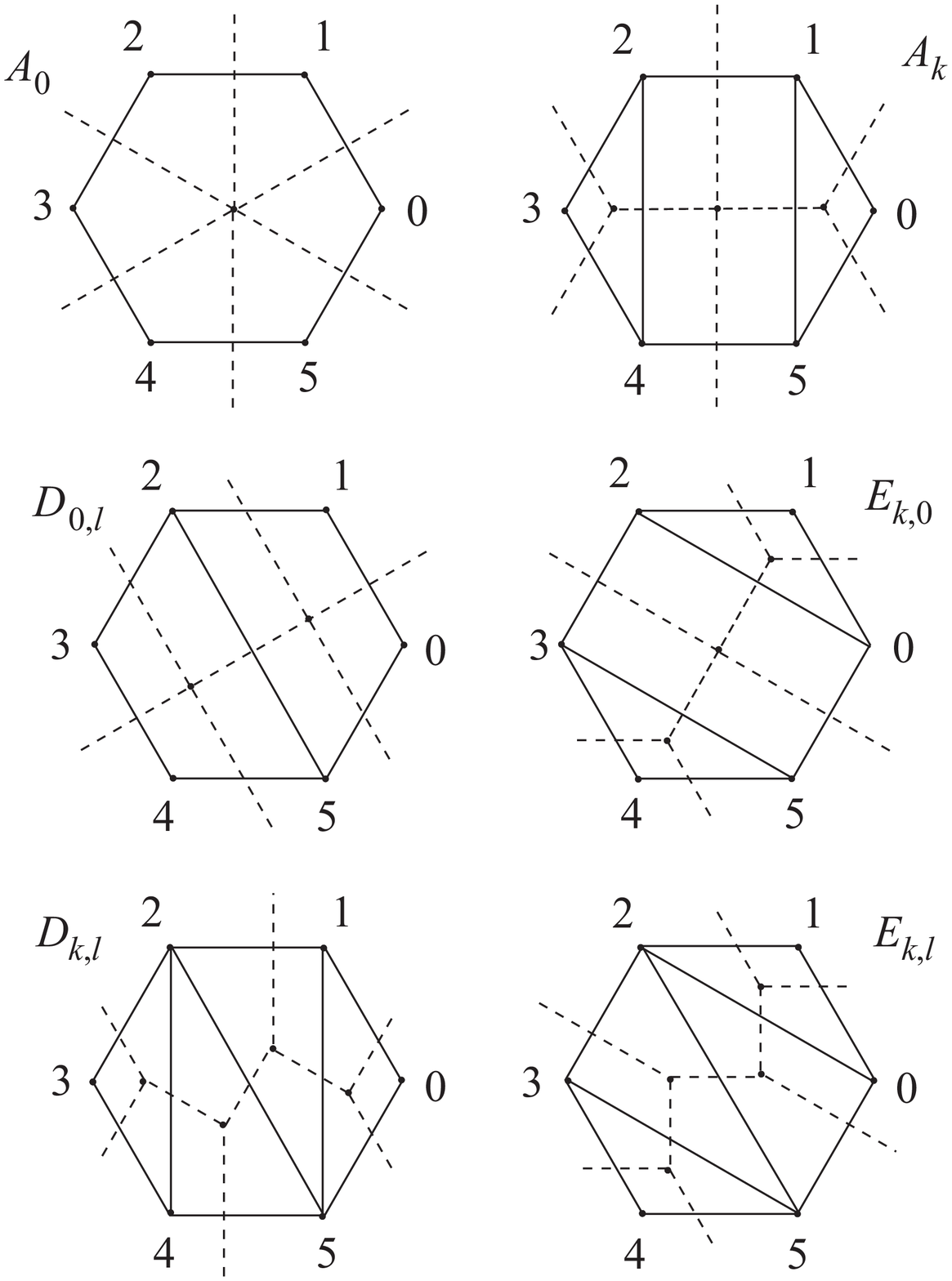}}
\nopagebreak
\vspace{.2in}\nopagebreak

Fig.~3. To the proof of Proposition 8.
\end{center}
The trees $A_k$ are symmetric with respect to complex conjugation.
The complex conjugates of the other trees are denoted by $\bar D_{k,l}$
and $\bar E_{k,l}$.
\vspace{.1in}

If the condition of Proposition 6 is not satisfied
(i.e., the cyclic order of the loops in $\Psi_0$ is different
from the cyclic order of the Stokes sectors) the graph $T$ may still be a tree when no three components
of its complement labeled by nonzero
 values are adjacent to any of its vertices.
This is the case for the trees
$E_{k,l}$ and their complex conjugates $\bar E_{k,l}$.
To distinguish them from the same
trees with the correct cyclic order of face labels
we denote them by $E'_{k,l}$ and $\bar E'_{k,l}$, respectively.
Examples of non-tree graphs $T$ are shown in Fig.~4.
\begin{center}
\epsfxsize=4.0in%
\centerline{\epsffile{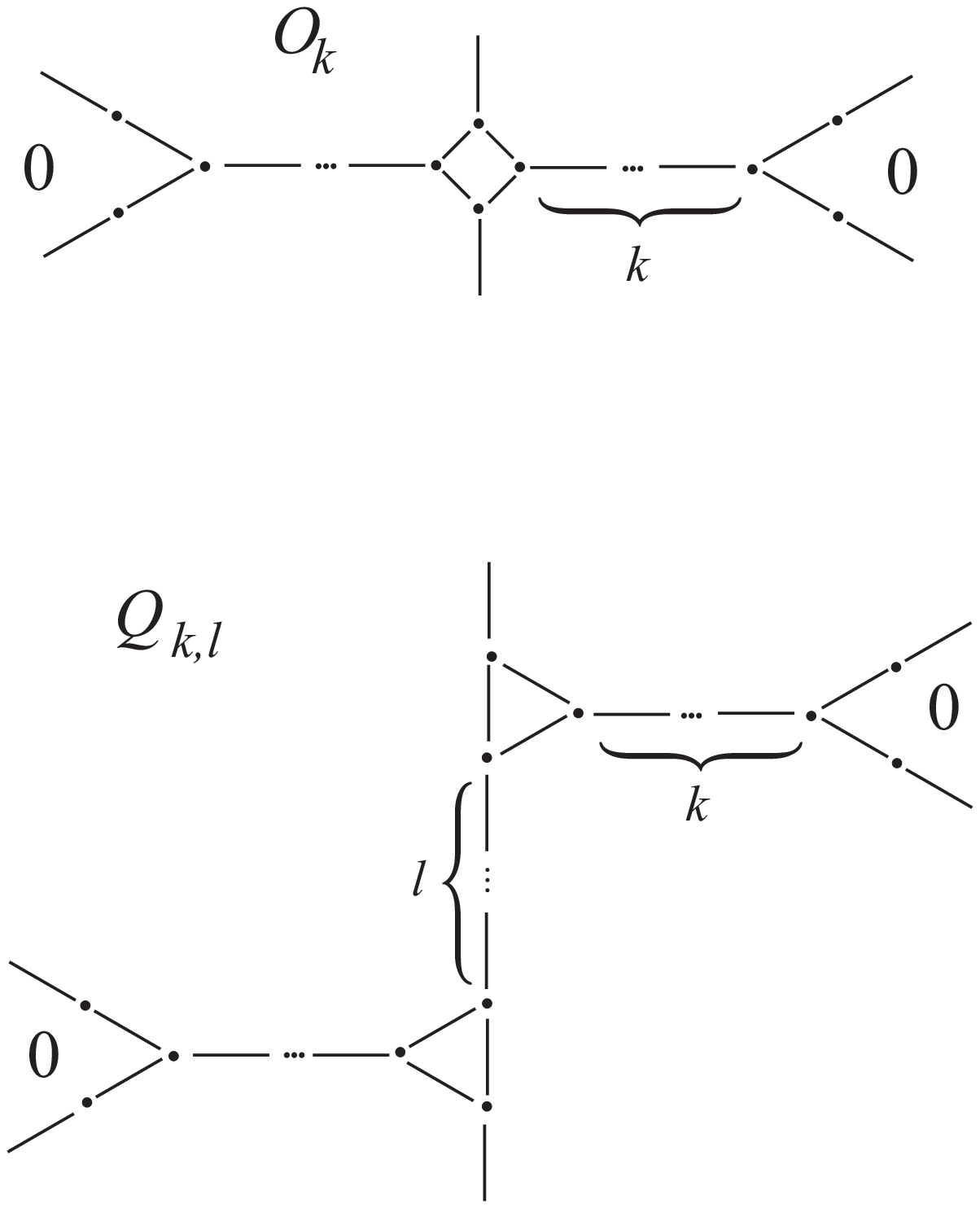}}
\nopagebreak
\vspace{.2in}\nopagebreak

Fig.~4. Non-tree examples of graphs $T$.
\end{center}
These graphs are $O_k$ and $Q_{k,l}$, with $k,l\ge 0$,
and the complex conjugates
$\bar Q_{k,l}$ of $Q_{k,l}$.
We will later show that these graphs $T$ can really occur.
Moreover, 
these are all possible cases of non-tree
graphs $T$ corresponding to cell decompositions $\Psi_f$
with the reverse cyclic order of non-zero labels, 
but we do not use this fact in the proof;
it will rather come as a consequence of our arguments.

\vspace{.2in}

\noindent
{\bf 5. Monodromy. Proof of Theorem 1 a).}
\vspace{.2in}

Let $\Xi$ be the space of all four-point sequences $\xi=\{c_1,\dots,c_4\}$ in $\bC_w$ such that
$c_3=-c_1,\,c_4=-c_2,\, c_\mu\ne 0,\infty,\pm c_\nu$ for $\mu\ne\nu$.
Let $\Xi_0$ be the quotient space of $\Xi$ with respect to multiplication by a nonzero constant.
A point of $\Xi_0$ can be represented by a sequence
$\{c,1,-c,-1\}$ such that $c\ne 0,\,\infty,\pm 1$.
Let $b=c^2\ne 0,\,\infty,\,1$. The fundamental group of $\bC\setminus\{0,\,\infty,\,1\}$
with a base point $-1$
is a free group with two generators $s_0$ and $s_\infty$ corresponding to the loops starting at $-1$,
going along the negative real axis towards $0$ (resp. $\infty$), going counterclockwise about $0$ (resp. $\infty$)
and returning to $-1$ along the negative real axis.
Each of these generators defines two paths
in the space $\Xi_0$ (denoted also by $s_0$ and $s_\infty$)
starting at $c=i$ (resp. $c=-i$) and ending at $c=-i$
(resp. $c=i$) (see Fig.~5).
\begin{center}
\epsfxsize=4.0in%
\centerline{\epsffile{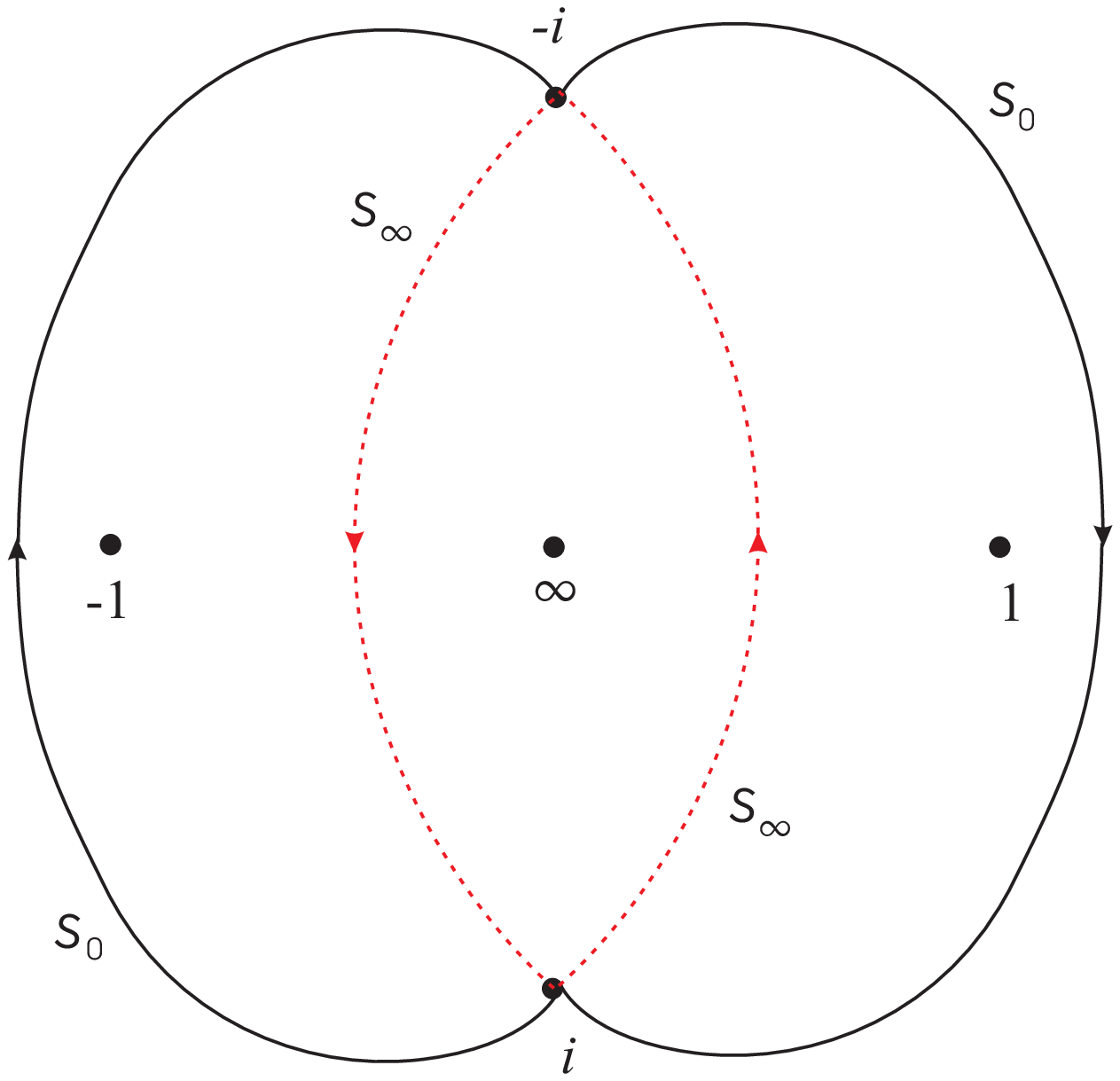}}
\nopagebreak
\vspace{.2in}\nopagebreak
Fig.~5. Paths $s_0$ and $s_{\infty}$.
\end{center}
The fundamental group of $\Xi_0$ with the base point $i$
is generated by $q_0=s_0^2,\; q_\infty=s_\infty^2,\; q_{-1}=(s_0 s_\infty)^{-1}$, and $q_{1}=(s_\infty s_0)^{-1}$,
with the relation $q_0 q_{1} q_{\infty} q_{-1}=1$.

Suppose that $f$ has nonzero asymptotic
values $c_1=i,\, c_2=1,\, c_3=-i,\, c_4=-1$.
Define four loops $\g_{i},\,\g_1,\,\g_{-i},\,\g_{-1}$
in $\C_w^\bullet$
by following either real or imaginary axis from
$\infty$ to one of the points $i,\, 1,\, -i,\, -1$, moving about that point counterclockwise,
and returning to $\infty$ along the same axis (Fig.~1).
These four loops generate the free group
 $\pi_1(\C_w^\bullet\setminus\{\pm i,\,\pm 1\})$.
We can assume that the cell
 decomposition $\Psi_0$ of $\C_w^\bullet$ defined
 by these four loops
is both centrally symmetric and invariant
 with respect to complex conjugation.
The cyclic order of the loops
 $\g_i,\,\g_1,\,\g_{-i},\,\g_{-1}$ at the point $\infty$ agrees
with the cyclic order of the corresponding Stokes sectors in $\C_z$.
According to propositions 6 and 7, the cell
 decomposition $\Psi_f$ of $\C_z$ can be defined by a tree $T$.

Each path $\{c(t),\, 1,\, -c(t), -1\}$ in $\Xi_0$
 starting at $c(0)=i$ 
defines a continuous
 deformation
\newline
 $\g_i(t),\,\g_1(t),\,\g_{-i}(t),\,\g_{-1}(t)$
 of the original four loops,
each of the deformed loops
 starting and ending at $\infty$ and
 avoiding $0,\,\infty,\,\pm 1,\,\pm c(t)$.
This deformation is unique up to isotopy.
We can choose the deformation so that
the corresponding cell decomposition $\Psi_0(t)$ of $\C_w^\bullet$
remains centrally symmetric.
For the paths corresponding
to $s_0$ and $s_\infty$, we have $c(1)=-i$.
Hence the loops
$\hat\g_{-i}=\g_i(1),\,\hat\g_1=\g_1(1),\,\hat\g_i=\g_{-i}(1),\,\hat\g_{-1}=\g_{-1}(1)$ belong to the same space
$\C_w^\bullet\setminus\{\pm i,\,\pm 1\}$
as the original loops $\g_{i},\,\g_1,\,\g_{-i},\,\g_{-1}$.
See Figs.~6, 7.
\begin{center}
\epsfxsize=4.5in%
\centerline{\epsffile{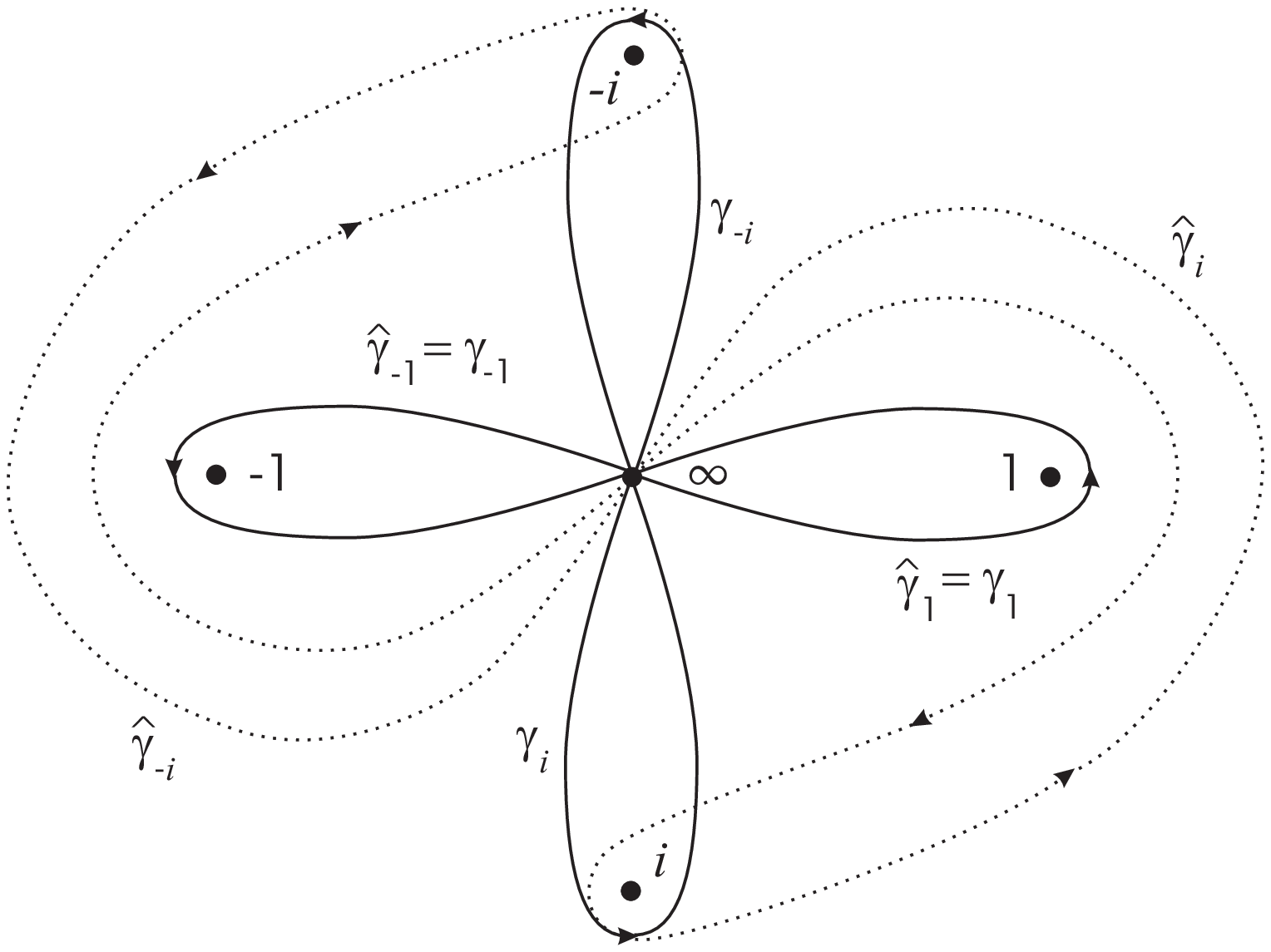}}
\nopagebreak
\vspace{.2in}\nopagebreak
Fig.~6. Action of $s_0$.
\end{center}
\begin{center}
\epsfxsize=3.5in%
\centerline{\epsffile{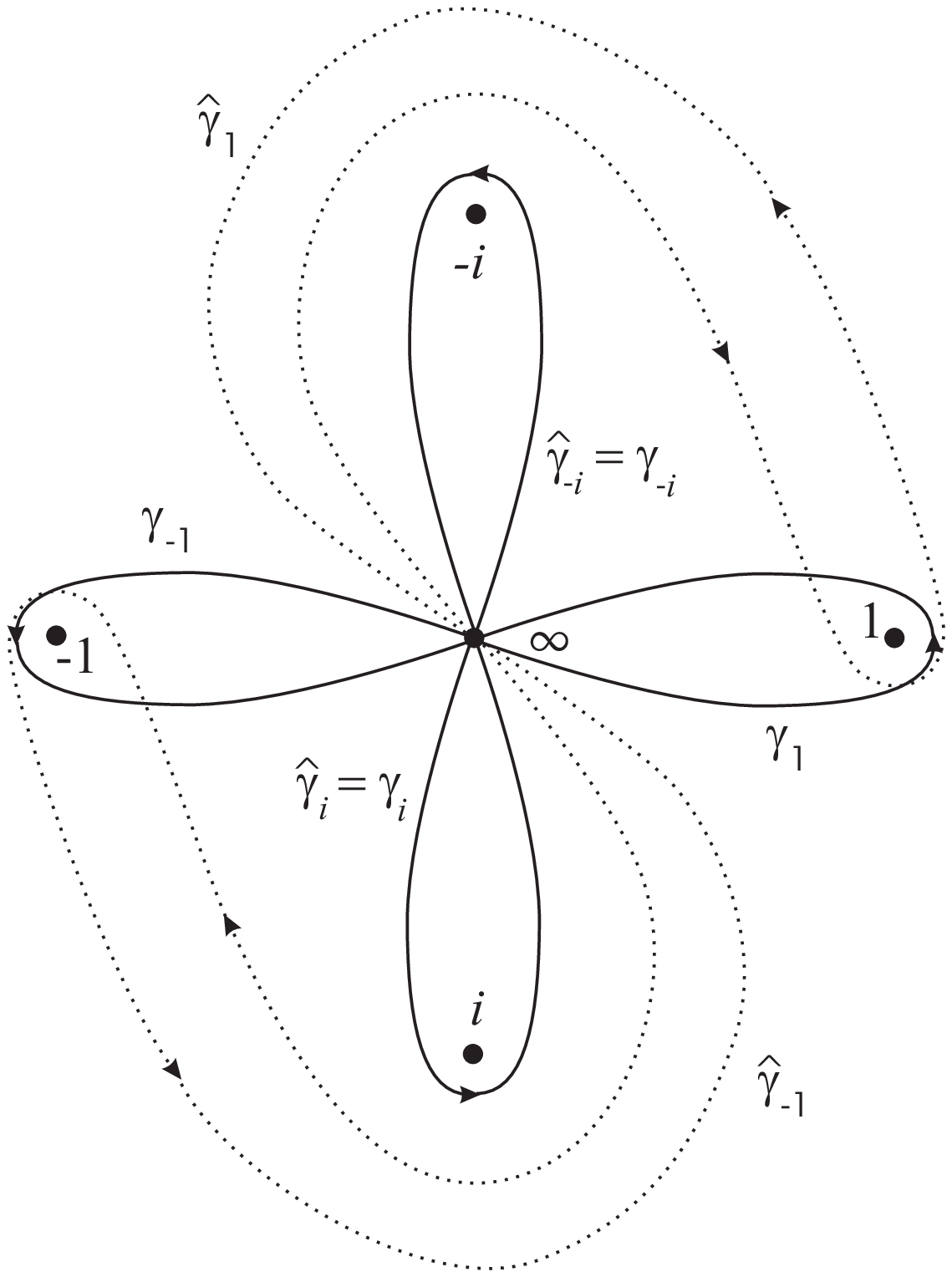}}
\nopagebreak
\vspace{.2in}\nopagebreak
Fig.~7. Action of $s_\infty$.
\end{center}
The paths in $\Xi_0$ corresponding to
$s_0$ and $s_\infty$ can be considered as elements of the braid group ${\mathcal B}_4$ on four
strands in $\C^\bullet_w$ (leaving $\pm 1$ fixed). 
From the classical formulas for the action of ${\mathcal B}_k$ on the fundamental group of the plane
without $k$ points \cite{LZ}, we have (see Fig.~6 for $s_0$
and Fig.~7 for $s_\infty$).

\begin{equation}\label{s0}
\hat\g_i=(\g_1)^{-1} \g_i \g_1,\; \hat\g_1=\g_1,\;\hat\g_{-i}=(\g_{-1})^{-1} \g_{-i} \g_{-1},\; \hat\g_{-1}=\g_{-1}\;\mbox{for}\;s_0;
\end{equation}
\begin{equation}\label{sinfty}
\hat\g_i=\g_i,\; \hat\g_1=(\g_{-i})^{-1} \g_1 \g_{-i},\; \hat\g_{-i}=\g_{-i},\; \hat\g_{-1}=(\g_i)^{-1} \g_{-1} \g_i,\;\mbox{for}\;s_\infty.
\end{equation}
Conversely, the original loops can be expressed as products of the new loops:
\begin{equation}\label{s0inv}
\g_i=\hat\g_1 \hat\g_i (\hat\g_1)^{-1},\; \g_1=\hat\g_1,\; \g_{-i}=\hat\g_{-1} \hat\g_{-i} (\hat\g_{-1})^{-1},\; \g_{-1}=\hat\g_{-1}\;\mbox{for}\;s_0;
\end{equation}
\begin{equation}\label{sinftyinv}
\g_i=\hat\g_i,\; \g_1=\hat\g_{-i} \hat\g_1 (\hat\g_{-i})^{-1},\; \g_{-i}=\hat\g_{-i},\; \g_{-1}=\hat\g_i \hat\g_{-1} (\hat\g_i)^{-1},\;\mbox{for}\;s_\infty.
\end{equation}

Let $f_t$ be the family of functions constructed in
 Proposition 6,
for the path
 $\{c(t),\,1,\,-c(t),-1\}$ in $\Xi_0$ with $c(0)=i$
and $c(1)=-i$, 
and let $\Psi_{f_t}=f_t^{-1}(\Psi_0(t))$ be the corresponding cell decompositions.
Note that the cell decomposition $\Psi_0(1)$ is defined by the loops
$\hat\g_{-i}=\g_i(1),\,\hat\g_1=\g_1(1),\,\hat\g_i=\g_{-i}(1),\,\hat\g_{-1}=\g_{-1}(1)$.
Then $f_1$ has the nonzero asymptotic values $-i,\,1,\,\,i,\,-1$.
The cell decomposition $\Psi_{f_1}$ is obtained
by a continuous deformation exchanging $i$ and $-i$
from the cell
decomposition $\Psi_{f_0}$. 
Accordingly, the directed graph without loops $\hat\G$ corresponding
to $\Psi_{f_1}$ is the same as the graph $\G$ corresponding to $\Psi_{f_0}$,
with the labels $i$ and $-i$ of its edges exchanged.

Let $\Psi'_{f_1}=f_1^{-1}(\Psi_0)$, and let $\G'$ be the corresponding
directed graph without loops.
Embedding of the graph $\G'$ in the plane $\C_z$ is determined,
up to an orientation-preserving homeomorphism of $\C_z$, by its
combinatorial structure and the cyclic order of its labeled directed edges at
each of its vertices of degree greater than $2$ \cite{LZ}.
Since this cyclic order agrees with the cyclic
order of directed edges of the cell decomposition $\Psi_0$ at its vertex $\infty$,
to define $\G'$ it is enough to specify
its combinatorial structure, i.e., for a vertex $v$ of $\G'$
(which can be identified with a vertex of $\hat\G$, since both graphs have $f_1^{-1}(\infty)$
as their vertices)
to determine whether an edge of $\G'$ labeled with one of $i,\, 1,\, -i,\, -1$
exits $v$, and if yes, which vertex of $\G'$ does it enter.

Two vertices $v$ and $v'$ of $\hat\G$ are connected by a directed edge labeled by $i$
(resp., $1,\, -i,\, -1$) if and only if the monodromy of $f_1^{-1}$ along the loop $\hat\g_i$
(resp., $\hat\g_1,\, \hat\g_{-i},\, \hat\g_{-1}$) maps $v$ to $v'\ne v$.
For $\G'$, the same holds for the monodromy of $f_1^{-1}$ along the loop $\g_i$
(resp., $\g_1,\, \g_{-i},\, \g_{-1}$).

Let us denote the corresponding
 monodromy transformations by 
\newline
$\sigma_i,\,\sigma_1,\,\sigma_{-i},\,\sigma_{-1}$, and by
$\hat\sigma_i,\,\hat\sigma_1,\,\hat\sigma_{-i},\,\hat\sigma_{-1}$, respectively.
Then the transformations $\hat\sigma$ can be determined from
the transformations $\sigma$ from the
relations (\ref{s0inv}) and (\ref{sinftyinv}) between the two sets of loops.
Note that the monodromy is anti-representation of the fundamental group,
i.e., the transformation corresponding to the product $\g\g'$ of two elements
of the fundamental group is $\sigma'\circ\sigma$, where $\sigma$ and $\sigma'$
are the monodromy transformations corresponding to
 $\g$ and $\g'$.

For $s_0$, the edges of $\G'$ labeled with $1$ and $-1$ are the same as the edges of $\hat\G$ with the same labels.
For each vertex $v$, the edge of $\G'$ labeled with $i$ that starts at $v$ ends at the vertex obtained from $v$
by moving along an edge of $\hat\G$ labeled with $1$ (or staying at $v$ if there is no such edge),
then along an edge of $\hat\G$ labeled with $i$, then
backward along
an edge of $\hat\G$ labeled with $1$. (There is no edge
when the last vertex coincides with $v$.)

The edge of $\G'$ labeled with $-i$ that starts at $v$ ends at the
vertex obtained from $v$ by moving along an edge of $\hat\G$ labeled with $-1$,
then along an edge of $\hat\G$ labeled with $-i$,
then backward along an edge of $\hat\G$ labeled with $-1$.

For $s_\infty$, the edges of $\G'$ labeled with $i$ and $-i$ are the same as the edges of $\hat\G$ with the same labels.
For each vertex $v$, the edge of $\G'$ labeled with $1$ that starts at $v$ ends at the vertex obtained from $v$
by moving along an edge of $\hat\G$ labeled with $-i$, then along an edge of $\hat\G$ labeled with $1$, then backward along
an edge of $\hat\G$ labeled with $-i$. The edge of $\G'$ labeled with $-1$ that starts at $v$ ends at the
vertex obtained from $v$ by moving along an edge of $\hat\G$ labeled with $i$, then along an edge of $\hat\G$ labeled with $-1$,
then backward along an edge of $\hat\G$ labeled with $i$.

Fig.~8 shows the action of $s_0$ for
 the graph $\G$ corresponding to the tree $T$ of type $A_1$.
The graph $\G'$ corresponds to an
 undirected graph $T'$ of type $Q_{1,0}$.
\begin{center}
\epsfxsize=5.0in%
\centerline{\epsffile{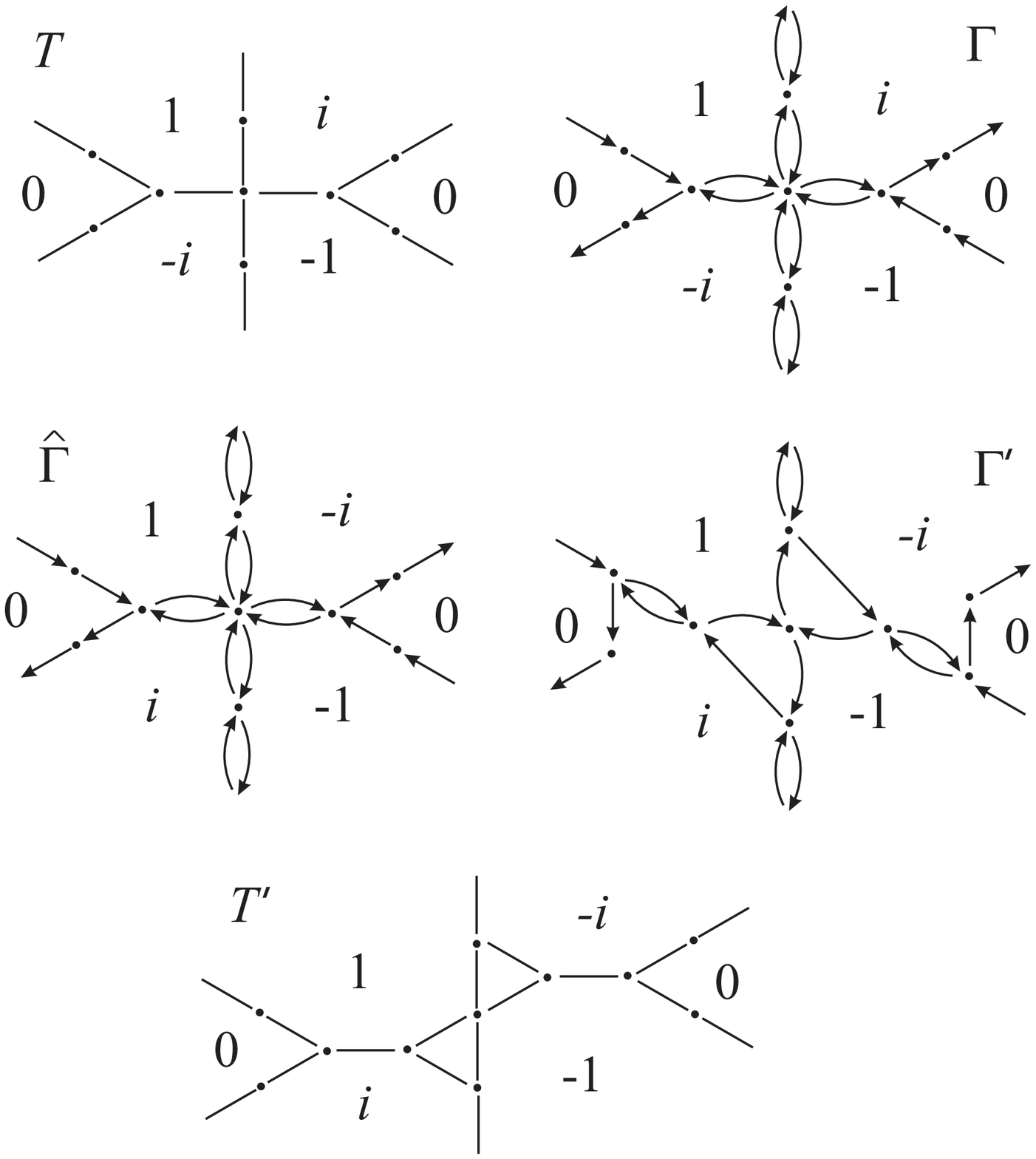}}
\nopagebreak
\vspace{.2in}\nopagebreak
Fig.~8. Action of $s_0$ transforms $A_1$ to $Q_{1,0}$.
\end{center}

Similarly, for the paths $(s_0)^{-1}$ and $(s_\infty)^{-1}$,
the transformations $\hat\sigma$ can be determined from
the transformations $\sigma$ from the relations (\ref{s0}) and (\ref{sinfty}).
However, we do not need this, since they can also be obtained from the symmetry with respect
to complex conjugation, sending a function $f(z)$ to $\bar f(\bar z)$.
Since the cell decomposition $\Psi_0$ in Fig.~1 is symmetric
with respect to complex conjugation,
the cell decomposition $\Psi_f$ and the corresponding graph $\G$
is exchanged its complex conjugate, $c$ with $1/\bar c$, 
$\alpha$ with $\bar\alpha$,
and the action of $s_0$ with the action of $s_\infty^{-1}$.

The action of $s_0$ and $s_\infty$ and their inverses
on all trees of Proposition~8
is summarized in the following tables (see also
Figs.~9, 10, where this action is represented
graphically).

$$
\begin{array}{ll}
s_0(A_k)=Q_{k,0},\;k\ge 0;&s_0(Q_{k,0})=A_{k+1},\; k\ge 0;\\
s_0(D_{k,l})=Q_{k,l},\;k,l\ge 0;&s_0(Q_{k,l})=D_{k+1,l},\; k,l\ge 0;\\
s_0(E_{k,l})=E'_{k,l},\;k\ge 1,\, l\ge 0;&s_0(E'_{k,l})=E_{k-1,l},\; k\ge 1,\, l\ge 0;\\
s_0(\bar D_{k,l})=\bar Q_{k,l-2},\;k\ge 0,\, l\ge 2;&s_0(\bar Q_{k,l})=\bar D_{k,l+2},\; k,l\ge 0;\\
s_0(\bar E_{k,l})=\bar E'_{k,l},\;k\ge 1,\, l\ge 0;&s_0(\bar E'_{k,l})=\bar E_{k,l},\; k\ge 1,\, l\ge 0;\\
s_0(\bar D_{k,1})=O_k,\; k\ge 0&s_0(O_k)=\bar D_{k,1},\;k\ge 0;\\
s_0(E'_{1,l})=D_{0,l},\;l\ge 0;&s_0(E'_{1,0})=A_0.
\end{array}
$$
\nopagebreak
\centerline{Table 1. The action of $s_0$.}
\vspace{.1in}
$$
\begin{array}{ll}
s_\infty(A_k)=\bar Q_{k-1,0},\;k\ge 1;&s_\infty(\bar Q_{k,0})=A_{k},\; k\ge 0;\\
s_\infty(D_{k,l})=Q_{k,l-2},\;k\ge 0,\,l\ge 2;&s_\infty(Q_{k,l})=D_{k,l+2},\; k,l\ge 0;\\
s_\infty(E_{k,l})=E'_{k,l},\;k\ge 1,\, l\ge 0;&s_\infty(E'_{k,l})=E_{k,l},\; k\ge 1,\, l\ge 0;\\
s_\infty(\bar D_{k,l})=\bar Q_{k-1,l},\;k\ge 1,\, l\ge 1;&s_\infty(\bar Q_{k,l})=\bar D_{k,l},\; k\ge 1,\, l\ge 1;\\
s_\infty(\bar E_{k,l})=\bar E'_{k+1,l},\;k\ge 1,\, l\ge 0;&s_\infty(\bar E'_{k,l})=\bar E_{k,l},\; k\ge 1,\, l\ge 0;\\
s_\infty(D_{k,1})=O_k,\; k\ge 0;&s_\infty(O_k)=D_{k,1},\;k\ge 0;\\
s_\infty(\bar D_{0,l})=\bar E'_{1,l},\;l\ge 0;&s_\infty(A_0)=\bar E'_{1,0}.
\end{array}
$$
\nopagebreak
\centerline{Table 2. The action of $s_\infty$.}
\newpage
\begin{center}
\epsfxsize=5.0in%
\centerline{\epsffile{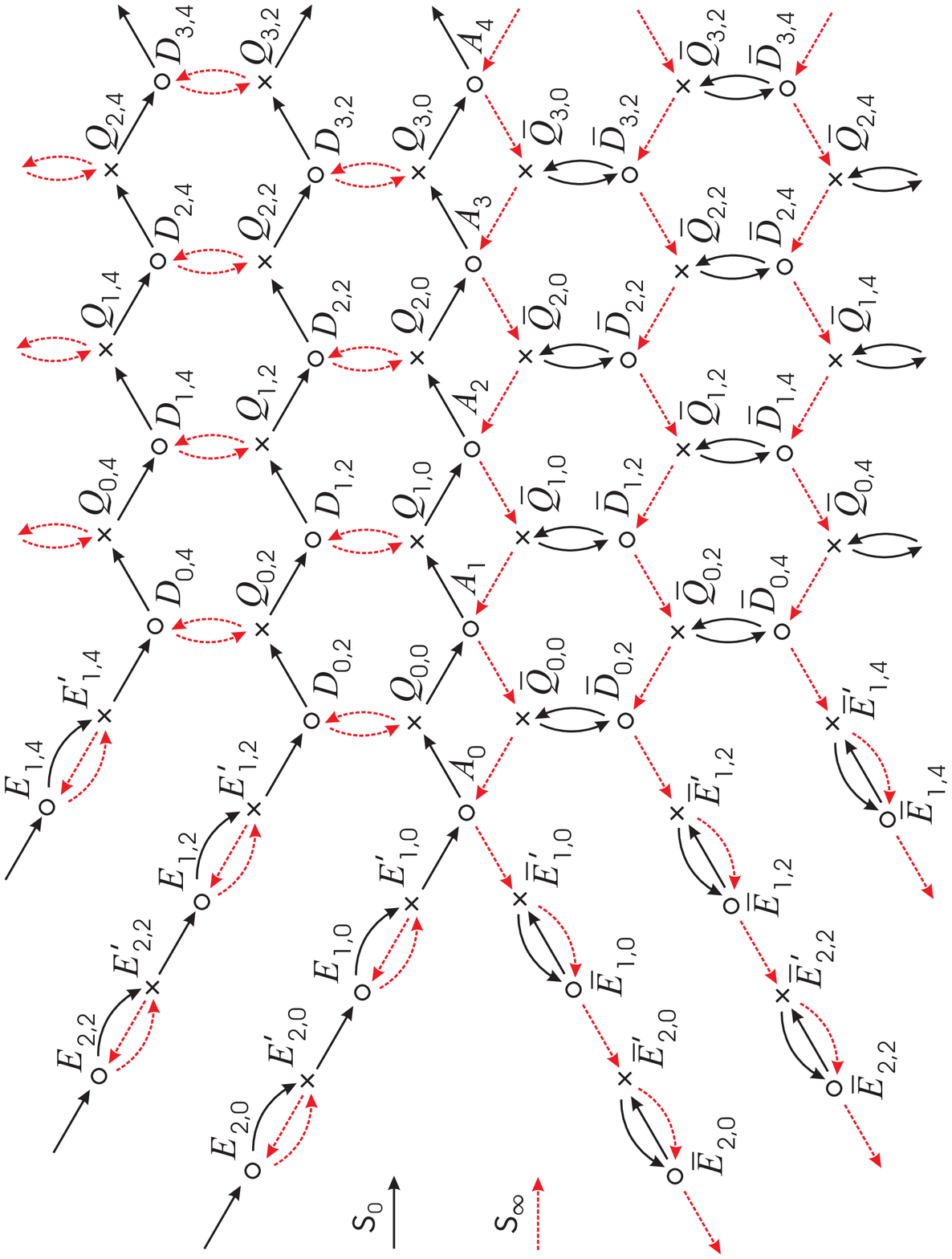}}
\nopagebreak
\vspace{.2in}\nopagebreak
Fig.~9. Monodromy action on the even part of $G$.
\end{center}
\vspace{.1in}

\begin{center}
\epsfxsize=5.0in%
\centerline{\epsffile{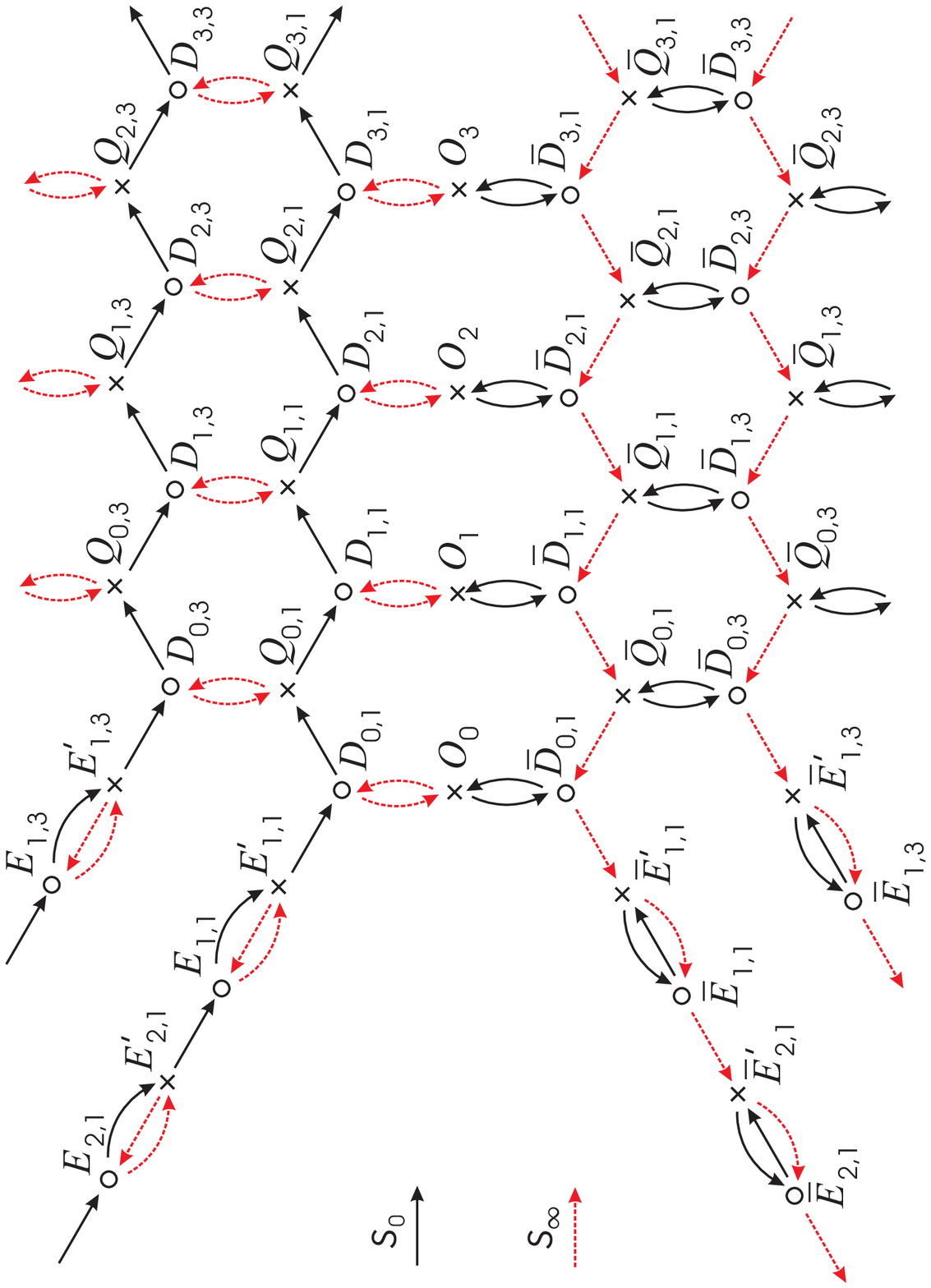}}
\nopagebreak
\vspace{.2in}\nopagebreak
Fig.~10. Monodromy action on the odd part of $G$.
\end{center}
\vspace{.1in}

\vspace{.1in}

\noindent
{\em Completion of the proof of Theorem 1 a).}
\vspace{.1in}

To prove that the Riemann surface $G$ has exactly two connected
components, corresponding to functions with a pole at the origin and the functions
with a zero at the origin, respectively, it is enough to show that,
for each point $c_0\in\bC$ over which the mapping
$W: G\to\bC$ is not ramified (i.e., $c_0\ne 0,\,\infty,\,\pm 1$) the monodromy of
$W^{-1}$ acts transitively on the fiber $W^{-1}(c_0)$ of $W$.
Let us choose a point $c_0=i$.
With the loops $\g_i,\,\g_1,\,\g_{-i},\,\g_{-1}$ defined above, the fiber of $W^{-1}(i)$
consists of the functions $f$ such that the cell decomposition $\Psi_f$ of $\C_z$
corresponds to one of the trees $A_k,\, D_{k,l},\, E_{k,l}, \bar D_{k,l},\, \bar E_{k,l}$.
From the tables for the action of $s_0$ and $s_\infty$, for the trees with a vertex
at the origin we have:

\noindent{\bf (i)} $A_k$ can be obtained from $A_0$ applying $(s_0)^{2k}$;

\noindent{\bf (ii)} $D_{k,2l}$ can be obtained from $A_k$ applying $(s_0 s_\infty)^l$;

\noindent{\bf (iii)} $E_{k,2l}$ can be obtained from $D_{0,2l}$ applying $(s_0)^{-2k}$;

\noindent{\bf (iv)} $E_{k,0}$ can be obtained from $A_0$ applying $(s_0)^{-2k}$;

\noindent{\bf (v)} The points corresponding to $\bar D_{k,2l}$ and $\bar E_{k,2l}$
can be obtained from $A_0$ combining the paths in (i-iv) with the complex
conjugation.

For the trees with no vertex at the origin we have:

\noindent{\bf (i)} $D_{k,1}$ can be obtained from $D_{0,1}$ applying $(s_0)^{2k}$;

\noindent{\bf (ii)} $D_{k,2l+1}$ can be obtained from $D_{k,1}$ applying $(s_0 s_\infty)^l$;

\noindent{\bf (iii)} $E_{k,2l+1}$ can be obtained from $D_{0,2l+1}$ applying $(s_0)^{-2k}$;

\noindent{\bf (iv)} $\bar D_{k,1}$ can be obtained from $D_{k,1}$ applying $s_\infty s_0$;

\noindent{\bf (v)} The points corresponding to $\bar D_{k,2l+1}$ and $\bar E_{k,2l+1}$
can be obtained from $D_{0,1}$ combining the paths in (i-iv) with the complex
conjugation.
\vspace{.1in}

\noindent{\em Eigenfunctions of self-adjoint operators.}
\vspace{.1in}

Suppose that $\alpha$ in (\ref{1})
is real, i.e., the problem is self-adjoint
and the eigenvalues are real.
Let $\lambda_0<\lambda_1<\dots$ be the eigenvalues of (\ref{1})
and $y_1(z),\;y_2(z),\dots$ the corresponding eigenfunctions.
Then $y_n(z)$ has $n$ real zeros (see \cite{Sibuya}).
Let $\lambda=\lambda_n$ be one of these eigenvalues.
Since $\alpha$ and $\lambda$ are real, the
function $f(z)$ defined in Section 2 is a real odd
meromorphic function.
Hence its nonzero asymptotic values satisfy
$c_2=-\bar c_1,\;c_3=-c_1,\;c_4=\bar c_1$.
These asymptotic values can be neither real nor pure imaginary
(see \cite{EGS}).
Let $a=c_1$, and let
$\Psi_a$ be a cell decomposition of $\bC_w$ 
defined similarly to the cell decomposition $\Psi_0$ in Fig.~1,
except the four loops of $\Psi_a$ contain the asymptotic values
$\pm a,\;\pm \bar a$ of $f$ (see Fig.~11).
\vspace{.1in}
\begin{center}
\epsfxsize=3.5in%
\centerline{\epsffile{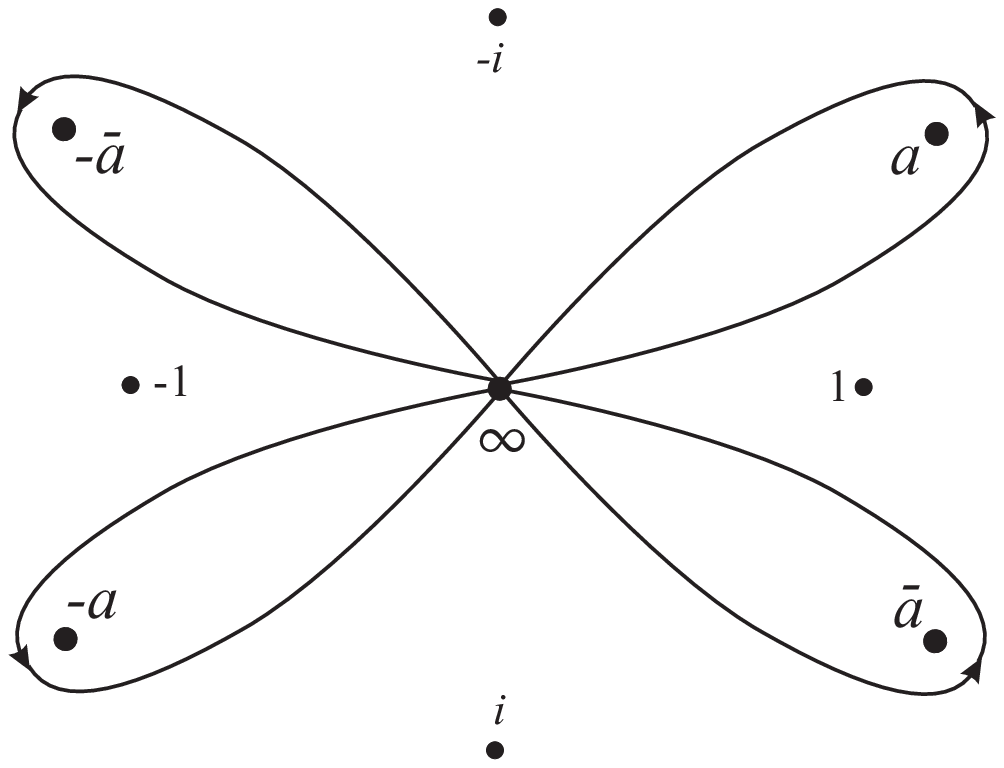}}
\nopagebreak
\vspace{.2in}\nopagebreak
Fig.~11. Cell decomposition for a real function.
\end{center}
\vspace{.1in}
We assume that $\Psi_a$ is centrally symmetric and
invariant under complex conjugation.
Then the cell decomposition $\Psi_f=f^{-1}(\Psi_a)$ of $\C_z$ is also
centrally symmetric and invariant under complex conjugation.
\vspace{.1in}

\noindent
{\bf Proposition 9.} {\em 
The cell decomposition $\Psi_f$ is of the type $A_k$ for $n=2k$, 
and of the type $O_k$ for $n=2k+1$.}
\vspace{.1in}

{\em Proof.}
Since $\Psi_f$ is invariant under complex
conjugation in ${\mathbf C}_z$, the corresponding graphs
$\Gamma$ and $T$ are symmetric under complex conjugation.

If $a$ is in the second or fourth quadrant
(as in Fig.~11) the cyclic order of the
nonzero asymptotic values of $f$ in $\C_w^\bullet$ is
consistent with the cyclic order of the corresponding
Stokes sectors $S_1,\;S_2,\;S_4,\;S_5$.
Thus conditions of Proposition 6 are satisfied, and
the graph $T$ should be one of the trees listed in Proposition 8.
Since the only such trees symmetric under complex conjugation are $A_k$,
we have $T=A_k$ for some $k$.
A real function $f$ with the cell decomposition $\Psi_f$
of the type $A_k$ has $2k$ real zeros (one zero in each two-gon
of $\Psi_f$ corresponding to an edge of $T$ on the real line).
Hence $n=2k$ in this case.

If $a$ is either in the first or third quadrant, the cyclic order
of the asymptotic values of $f$ is opposite to the cyclic
order of the corresponding Stokes sectors.
In this case, operations $s'_0$ and $s'_\infty$ similar to $s_0$ and
$s_\infty$ (Figs.~6 and 7) can be applied to reverse the cyclic order.
More precisely, for $a=\pm e^{i\pi/4}$, the function
$g(z)=e^{i\pi/4}f(z)$ has the asymptotic values $\pm 1,\;\pm i$
and the cell decomposition $\Psi_g=g^{-1}(\Psi_0)$, where
$\Psi_0$ is the cell decomposition in Fig.~1,
coincides with $\Psi_f$.
The action of $s'_0$ and $s'_\infty$ on $\Psi_f$ is defined
as the action of $s_0$ and $s_\infty$ on $\Psi_g$.
For any $a$ in the first or third quadrant, one can 
deform $\pm a$ to $\pm e^{i\pi/4}$ without crossing the real and
imaginary exes, apply the two operations, and deform $\pm a$ back to
initial values.
Thus the action of $s'_0$ and $s'_\infty$ on $\Psi_f$
is the same as the action of $s_0$ and $s_\infty$
described in Tables 1 and 2.
In particular, the result of this action is a cell
decomposition satisfying the condition of Proposition
6, hence having the type of one of the trees listed
in Proposition 8. Accordingly, the cell decomposition
$\Psi_f$ itself should correspond to a graph $T$
obtained from one of these trees 
by the operations $s_0^{-1}$ and $s_\infty^{-1}$.
From Tables 1 and 2, all such graphs are of the types
$O_k,\; Q_{k,l},\; E'_{k,l},\; \bar Q_{k,l},\;\bar E'_{k,l}$.
The only graphs in this list symmetric under complex conjugation
are $O_k$. A real function $f$ with the cell decomposition 
$\Psi_f$ of the type $O_k$ has $2k+1$ real zeros
(one zero at the origin and one inside each two-gon
of $\Psi_f$ corresponding to an edge of $T$ on the real line).
Hence $n=2k+1$ in this case.
\vspace{.2in}

\noindent
{\bf 6. Other potentials}
\vspace{.2in}

The following three one-parametric families
of potentials can be treated with the same method.
The details will appear elsewhere.
\vspace{.1in}

1. PT-symmetric cubic \cite{DT}. A differential
equation $-y^{\prime\prime}+P(z)y=\lambda y$, with a
general cubic
polynomial $P$,
by an affine change of the independent variable
and a shift of $\lambda$ can be brought to the form
\begin{equation}
\label{cubic}
-y^{\prime\prime}+(iz^3+i\alpha z)y=\lambda y.
\end{equation}
We impose the boundary condition
\begin{equation}
\label{bc}
y(z)\to 0,\quad \mbox{as}\quad y\in\R,\quad y\to\pm\infty.
\end{equation}
This problem is not self-adjoint 
but for real $\alpha$ it has the
so-called PT-symmetry property. It is known
\cite{DDT,Shin7} that for $\alpha\geq 0$ the spectrum 
is real.
\vspace{.1in}

\noindent
{\bf Theorem 3.} {\em Let $Z_3\in\C^2$ be
the set of all pairs
$(\alpha,\lambda)$ such that the problem (\ref{cubic}),
(\ref{bc}) has a non-trivial solution.
Then $Z_3$ is an irreducible non-singular curve.}
\vspace{.1in}

2. Quasi-exactly solvable sextic family \cite{TU,Ushv}.
Consider the equation
\begin{equation}
\label{sextic}
-y^{\prime\prime}+(z^6+2\alpha z^4+\{\alpha^2-(4m+2p+3)\}z^2)y=
\lambda y,
\end{equation}
with the same boundary condition (\ref{bc}).
This problem is self-adjoint for real $\alpha$.
It was shown by Turbiner
and Ushveridze that for real $\alpha$ this problem has 
exactly $m+1$ linearly independent ``elementary''
eigenfunctions of the form $Qe^T$ with polynomials
$Q$ and $T$. The degree of $Q$ is $2m+p$, so the eigenfunction
has $2m+p$ zeros in the complex plane. If $p=0$ then
these elementary eigenfunctions correspond to 
the first $m+1$ even-numbered eigenvalues,
and if $p=1$ to the first $m+1$ odd-numbered eigenvalues. 
\vspace{.1in}

\noindent
{\bf Theorem 4.} {\em Let $m$ be a non-negative
integer and $p\in\{0,1\}$. Let
$Z_{6,m,p}\in\C^2$ be the set of
all pairs $(\alpha,\lambda)$ such that $\lambda$ is
an eigenvalue of the problem (\ref{sextic}), (\ref{bc})
corresponding to an elementary eigenfunction.
Then $Z_{6,m,p}$ is a non-singular irreducible
curve. }
\vspace{.1in}

3. Quasi-exactly solvable PT-symmetric
quartic family \cite{Bender1}. Consider the equation
\begin {equation}
\label{qq}
-y^{\prime\prime}+(-z^4-2\alpha z^2-2imz)y=\lambda y.
\end{equation}
Here the boundary condition is 
\begin{equation}
\label{qbc}
y(re^{i\theta})\to 0,\quad\mbox{as}\quad r\to\infty,\quad
\theta\in \{-\pi/6,-\pi+\pi/6\}.
\end{equation}
Similarly to the previous example, this problem
has $m$ elementary eigenfunctions.
\vspace{.1in}

\noindent
{\bf Theorem 5.} {\em Let $m$ be a non-negative integer,
and $Z_{4,m}\subset\C^2$ be the set of pairs
$(\alpha,\lambda)$ such that $\lambda$ is an 
eigenvalue of the problem (\ref{qq}), (\ref{qbc}),
with an elementary    eigenfunction.
Then $Z_{4,m}$ is a smooth irreducible curve.}

{\em Purdue University

West Lafayette IN 47907

USA}
\end{document}